\documentclass[a4paper,11pt]{article}

\usepackage{jheppub}
\usepackage[T1]{fontenc}
\usepackage{lineno}
\usepackage{booktabs}
\usepackage{multicol}
\usepackage{tabularx}
\usepackage{makecell}
\newcolumntype{D}{>{\centering \hsize=.2\hsize \arraybackslash}X}
\newcolumntype{C}{>{\centering \hsize=.4\hsize \arraybackslash}X}

\newcolumntype{?}{!{\vrule width 1.2pt}}

\usepackage{amsmath}
\usepackage{slashed}

\usepackage{graphicx}
\usepackage{subcaption}
\usepackage{tikz}
\usepackage[compat=1.1.0]{tikz-feynman}

\usetikzlibrary{arrows.meta}
\usetikzlibrary{decorations.markings}
\usetikzlibrary{decorations.pathmorphing}
\usetikzlibrary{shapes.geometric}
\pgfdeclarelayer{nodelayer}
\pgfdeclarelayer{edgelayer}
\pgfsetlayers{main,edgelayer,nodelayer}
\tikzstyle{none}=[]

\tikzstyle{blank}=[fill=white, shape=circle, draw=white, inner sep=0.8pt]
\tikzstyle{dot}=[fill=black, shape=circle, draw=black, inner sep=0.8pt]
\tikzstyle{fat}=[fill=blue, shape=circle, draw, line width=1pt, inner sep=0.8pt, minimum size=4mm]
\tikzstyle{blob}=[fill=gray, shape=circle, draw=black, line width=1pt, inner sep=0.8pt, minimum size=1cm]
\tikzstyle{star}=[fill=black, shape=star, inner sep=0.8pt, minimum size=5mm]
\tikzstyle{arrow}=[{-{Classical TikZ Rightarrow[length=2mm,width=1.5mm]}}, draw={rgb,255: red,61; green,171; blue,83}, line width=1pt, preaction={{draw=white,line width=2pt}}, line cap=rect]

\tikzstyle{gluoncoil}=[-, decorate, decoration={coil,aspect=1.4,segment length=2.5mm}]
\tikzstyle{fermion}=[-, draw={rgb,255: red,176; green,36; blue,39}, line width=1pt, preaction={{draw=white,line width=2pt}}, line cap=rect, postaction=decorate, decoration={markings,mark=at position .60 with {\arrow{Stealth[round,width=5pt]}}}]
\tikzstyle{scalar}=[-, line width=1pt, dashed, draw]
\tikzstyle{fermionarrow}=[-, postaction=decorate, decoration={markings,mark=at position .60 with {\arrow{Stealth[round,width=5pt]}}}]
\tikzstyle{ct scalar}=[-, line width=1pt, dashed, draw={rgb,255: red,102; green,102; blue,102}, postaction=decorate, decoration={markings,mark=at position .50 with {\node[style=star,minimum size=5mm]{};}}]
\tikzstyle{ct fermion}=[-, line width=1pt, preaction={{draw=white,line width=2pt}}, line cap=rect, postaction=decorate, decoration={markings,mark=at position 0.25 with {\arrow{Stealth[round,width=5pt]}}, mark=at position 0.75 with {\arrow{Stealth[round,width=5pt]}}, mark=at position .50 with {\node[style=star,minimum size=5mm]{};}}]

\newcommand{\pysecdec}{py{\textsc{SecDec}}}

\title{\boldmath Electroweak corrections to Higgs boson pair production: The top-Yukawa and self-coupling contributions}

\author[a]{G. Heinrich,}
\author[b]{S. P. Jones,}
\author[a]{M. Kerner,}
\author[b]{T. W. Stone,}
\author[a]{A. Vestner}
\affiliation[a]{Institute for Theoretical Physics, Karlsruhe Institute of Technology (KIT), 76128 Karlsruhe, Germany}
\affiliation[b]{Institute for Particle Physics Phenomenology, Durham University, Durham DH1 3LE, UK}

\emailAdd{gudrun.heinrich@kit.edu} 
\emailAdd{stephen.jones@durham.ac.uk}
\emailAdd{matthias.kerner@kit.edu}
\emailAdd{thomas.w.stone@durham.ac.uk}
\emailAdd{augustin.vestner@kit.edu}


\preprint{{\small  KA-TP-11-2024, IPPP/24/43, P3H-24-046}}

\abstract{
 We present results for the Yukawa-enhanced and Higgs self-coupling type electroweak corrections to di-Higgs production in gluon fusion. The calculation of the corresponding four-scale, two-loop amplitude is carried out retaining the exact symbolic dependence on all masses and scales during the reduction to master integrals. The resulting integrals are then evaluated at high precision using both the series expansion of the differential equations and sector decomposition. Differential cross sections for the di-Higgs invariant mass and the transverse momentum of a Higgs boson are shown, where we find that the corrections are most pronounced at low invariant mass and transverse momentum.}

\begin{document}
\maketitle
\flushbottom
\allowdisplaybreaks

\section{Introduction}
\label{sec:introduction}
Among the main goals of the current and future runs of the CERN Large Hadron Collider (LHC) is tightening the constraints on the trilinear Higgs boson self-coupling, $\lambda_{HHH}$. 
Measurements of the Higgs boson pair cross section at ATLAS currently set an upper limit of $\mu_{HH} < 2.4$ at 95\% confidence level, giving a bound on the self-coupling modifier, $\kappa_{\lambda}=\lambda_{HHH}/\lambda^{SM}_{HHH}$, of $-1.4 \nolinebreak <\nolinebreak \kappa_{\lambda}\nolinebreak < \nolinebreak 6.1$~\cite{ATLAS:2022jtk,ATLAS:2024ish} ($-0.4<\kappa_{\lambda}<6.3$ when combining both single and double Higgs production),  while measurements at CMS place a limit of $\mu_{HH} < 3.4$, giving $-1.24\nolinebreak<\nolinebreak\kappa_{\lambda}\nolinebreak<\nolinebreak 6.49$~\cite{CMS:2022dwd} ($-1.2<\kappa_{\lambda}<7.5$ when combining both single and double Higgs production~\cite{CMS-PAS-HIG-23-006}). 
The high-luminosity LHC run is expected to shrink this constraint to $0.1 \nolinebreak<\nolinebreak \kappa_{\lambda} \nolinebreak<\nolinebreak 2.3$ \cite{Micco_2020}, thereby ruling out the scenario of $\kappa_\lambda=0$ where the Higgs boson does not couple to itself via a cubic coupling. 
Higgs boson pair production in gluon fusion is the prime process to consider in order to constrain the trilinear coupling because $\lambda_{HHH}$ already enters at the leading order (LO) and the gluon fusion production channel is dominant at the LHC. Most physics models beyond the Standard Model (SM) predict modified Higgs boson self-couplings, particularly those where electroweak symmetry breaking occurred through a first-order phase transition, which is a prerequisite for generating the observed baryon asymmetry. 
Therefore, it is crucial to have precise predictions for this process within the SM, such that potential discrepancies between data and theory can be clearly identified as signposts of new physics.

The LO cross section for the process $gg\to HH$ has been calculated in Refs.~\cite{Eboli:1987dy,Glover:1987nx}, NLO QCD corrections including the full top-quark mass dependence are also
available~\cite{Borowka:2016ehy,Borowka:2016ypz,Baglio:2018lrj,Davies:2019dfy,Baglio:2020ini}, increasing the total cross section by about 60\%. 
The real corrections, entering the NLO QCD cross section, have recently been obtained in a compact analytic form~\cite{Campbell:2024tqg}.
NLO matching to parton showers has been performed in~\cite{Heinrich:2017kxx,Jones:2017giv,Heinrich:2019bkc,Bagnaschi:2023rbx}, later also including anomalous couplings within an Effective Field Theory (EFT) framework~\cite{Heinrich:2020ckp,Heinrich:2022idm,Heinrich:2023rsd}.
QCD corrections beyond NLO have been calculated in the heavy-top-limit~\cite{Deflorian:2018eng,deFlorian:2016uhr,Grigo:2015dia}, or in a combination of large-$m_t$ and high-energy expansions~\cite{Davies:2021kex}.
Partial three-loop results also have been obtained recently~\cite{Davies:2023obx,Davies:2024znp}.
The full NLO QCD corrections have been included in calculations where even higher orders have been evaluated, e.g. including the top mass dependence in the real corrections at NNLO~\cite{Grazzini:2018bsd}, or N$^3$LO corrections~\cite{Chen:2019lzz,Chen:2019fhs} and N$^3$LO+N$^3$LL corrections~\cite{AH:2022elh} in the heavy-top-limit. The N$^3$LO results have a residual scale uncertainty of about 3\%, therefore other uncertainties, such as missing electroweak (EW) corrections, become an important part in the uncertainty budget.
Currently, the top-mass renormalisation scheme uncertainties are the largest uncertainties for this process~\cite{Baglio:2020wgt,Bagnaschi:2023rbx}, they are estimated to be of the order of 20\%.
However, the electroweak corrections also introduce a renormalisation scheme dependence, and its interplay with the scheme dependence of the QCD corrections is currently unknown.
Furthermore, it is well-known that EW corrections can significantly affect the shape of kinematic distributions.
For example, the EW corrections to single Higgs boson production are of the order of +5\% for $m_H=125$\,GeV, dominated by the light fermion contributions, but, for larger values of $m_H$, the corrections become negative and the light quark contribution ceases to dominate the correction~\cite{Aglietti:2004nj,Degrassi:2004mx,Actis:2008ug}.
First partial NLO EW corrections to Higgs boson pair production have been calculated in Refs.~\cite{Borowka:2018pxx,Muhlleitner:2022ijf,Davies:2022ram}, the full NLO EW corrections in the large top-quark mass expansion up to $1/m_t^8$ have been calculated in Ref.~\cite{Davies:2023npk}. 
The possibility to constrain the quartic Higgs boson coupling indirectly through EW corrections to Higgs boson pair production has been explored in Refs.~\cite{Borowka:2018pxx,Bizon:2018syu,Bizon:2024juq}.
Total and differential cross sections including the full NLO EW corrections have been presented in Ref.~\cite{Bi:2023bnq}, finding a decrease by $-4$\% of the total cross section after inclusion of the NLO EW corrections.

As the first order EW corrections to double Higgs production factorise from the NLO QCD corrections to this process, mixed QCD-EW corrections would only play a role at even higher orders. The latter are relevant for single Higgs production, where the experimental uncertainties are very small; contributions to these mixed corrections have been calculated in Refs.~\cite{Anastasiou:2008tj,Bonetti:2016brm,Bonetti:2017ovy,Anastasiou:2018adr,Bonetti:2018ukf,Bonetti:2020hqh,Becchetti:2020wof,Bonetti:2022lrk}.

In this paper, we calculate the electroweak corrections to the process $gg\to HH$ in the scalar sector, i.e. the corrections which are Yukawa-enhanced or are of Higgs self-coupling type (with the quartic coupling $\lambda_{HHHH}$ also now entering at NLO), while corrections due to the exchange of virtual electroweak gauge bosons are not included.
The calculation involves four-point, two-loop integrals with up to two mass scales ($m_t$, $m_H$) and two independent Mandelstam variables ($s$, $t$), which we retain fully symbolically in our amplitude. The master integrals are evaluated in two ways: with the method of sector decomposition using {\sc pySecDec}~\cite{Borowka:2017idc,Borowka:2018goh,Heinrich:2021dbf,Heinrich:2023til} and by solving differential equations via series expansions using {\sc DiffExp}~\cite{Hidding:2020ytt,Moriello:2019yhu}.

The outline of this article is as follows: in Section~\ref{sec:calculation}, we give details of the calculation, describing the projection onto form factors, the reduction to master integrals and their evaluation; the UV renormalisation of the amplitude is also described in detail.
In Section~\ref{sec:results} we provide values for the bare and renormalised amplitude at a selected phase-space point and present our results for the Higgs boson pair invariant mass and Higgs boson transverse momentum distributions in addition to the impact of these corrections on the total cross section.
Our conclusions are presented in Section~\ref{sec:conclusions}.

\section{Calculation}
\label{sec:calculation}
In this section, we describe the details of our calculation of the NLO electroweak corrections to Higgs boson pair production including only the top-Yukawa and Higgs boson self-coupling contributions.
We start by specifying the parts of the SM Lagrangian relevant for computing these corrections in Section~\ref{sec:model}, followed by a detailed description of the amplitude structure for $gg\to HH$ in Section~\ref{sec:amplitude}. 
The remaining sections give details of our computational setup, starting from diagram generation in Section~\ref{sec:setup}, continuing with the reduction to master integrals in Section~\ref{sec:reduction}, and closing with the master integral evaluation in Section~\ref{sec:integrals}. 
In Section~\ref{sec:renormalisation}, we describe the renormalisation of our amplitudes. For a review of the standard methods for the computation and renormalisation of one-loop electroweak corrections in the Standard Model see Refs.~\cite{Denner:1991kt,Denner:2019vbn}.

\subsection{Lagrangian \& Input-Parameter Scheme}
\label{sec:model}
To precisely define the corrections we wish to compute -- i.e. only those induced by the Yukawa coupling and Higgs self-couplings -- and to derive their renormalisation, we do not start from the general SM Lagrangian. 
Instead, we start from a more accessible subset corresponding to a Yukawa model with only one up-type fermion (the top quark) and one scalar field (the Higgs boson). 
Indeed, employing a series of simplifications, we can see it truly is a subset of the SM: firstly, we remove the Yang-Mills part for the electroweak gauge bosons so that they only appear in the covariant derivative.
Additionally, all leptons, light quarks and the bottom quark are dropped since their coupling to the Higgs field is negligibly small compared to that of the top quark. Prior to electroweak symmetry breaking (EWSB), this leads to the bare Lagrangian
\begin{equation}\begin{split}
	\mathcal{L}_0=&-\frac14 \mathcal G_{0,\mu\nu}\mathcal G_0^{\mu\nu} + (D_\mu\Phi_0)^\dagger(D^\mu\Phi_0) + \mu_{0}^2\Phi_0^\dagger\Phi_0 + \frac{\lambda_0}{4}(\Phi_0^\dagger\Phi_0)^2\\
    &+ i \bar Q_{L,0} \slashed D Q_{L,0} + i \bar t_{R,0}\slashed D t_{R,0} - (y_{t,0}\bar Q_{L,0}\Phi_0^ct_{R,0} + \mathrm{h.c.}),
\end{split}\end{equation}
with
\begin{equation}
	Q_{L,0}=\begin{pmatrix}
		t_{L,0}\\0
	\end{pmatrix},
\end{equation}
and the gluon field strength tensor $\mathcal G_0^{\mu\nu}$.
Taking the gaugeless limit for the EW sector corresponds to the limit $ (g,g')\rightarrow (0,0) $, which removes the electroweak gauge bosons (as well as their associated ghost fields) entirely, such that the covariant derivatives have the form
\begin{equation}
	D_\mu = \partial_\mu - i g_{s,0} G_{0,\mu}^a t^a\;
\end{equation}
where $ t^a $ are the generators of SU(3)$_\mathrm{colour}$ and $G_{0,\mu}^a$ are the gluon fields. 
Through EWSB the Higgs field $ \Phi_0 $ acquires a vacuum expectation value (vev) $v_0$. 
Expanding the Higgs field around its vev and using unitary gauge to decouple the Goldstone bosons we obtain the Lagrangian
\begin{align}\begin{split}
	\mathcal{L}_0=&-\frac14 \mathcal G_{0,\mu\nu}\mathcal G_0^{\mu\nu} + \frac{1}{2}(\partial_\mu H_0)^\dagger(\partial^\mu H_0) + \frac{\mu_0^2}{2} (v_0 + H_0)^2 + \frac{\lambda}{16} (v_0 + H_0)^4\\
    &+ i \bar t_0 \slashed D t_0 - y_{t,0} \frac{v_0+H_0}{\sqrt{2}}\bar t_0t_0 + \mathrm{constant}\nonumber\end{split}\\
	\begin{split}=&-\frac14 \mathcal G_{0,\mu\nu}\mathcal G_0^{\mu\nu} + \frac{1}{2} (\partial_\mu H_0)^\dagger(\partial^\mu H_0) - \frac{m_{H,0}^2}{2}H_0^2 - \frac{m_{H,0}^2}{2v_0} H_0^3 - \frac{m_{H,0}^2}{8v_0^2} H_0^4\\
    &+ i \bar t_0 \slashed D t_0 - m_{t,0} \bar t_0t_0 - \frac{m_{t,0}}{v_0} H_0\bar t_0t_0 + \mathrm{constant}\label{eq:LLO}\end{split}
\end{align}
with the identifications,
\begin{align}
	m_{H,0}^2&=2\mu_0^2\; , \;
	m_{t,0}=\frac{y_{t,0}v_0}{\sqrt{2}}\;\mbox{ and }\;
	v_0^2=-\frac{2m_{H,0}^2}{\lambda_0}~.\label{eq:mu_vev_mh_relation}
\end{align}
The constant term in the Lagrangian is neglected from now on as it does not contribute to observables. 
For later convenience, we also introduce the labels $g_{t,0}, g_{3,0}, g_{4,0}$ (and $g_{t}, g_{3}, g_{4}$) for the bare (and renormalised) top-Yukawa ($H\overline{t}t$) coupling, trilinear Higgs ($H^3$) self-coupling and quartic Higgs ($H^4$) self-coupling, respectively.
In the SM and in our Yukawa model, they are related to the top-quark mass, Higgs boson mass and vev via
\begin{align}
&g_{t,0} \equiv \frac{m_{t,0}}{v_0}~,& 
&g_{3,0} \equiv \frac{3m_{H,0}^2}{v_0}~,&
&g_{4,0} \equiv \frac{3m_{H,0}^2}{v_0^2}~.&
\label{eq:SMcouplings}
\end{align}
We present the set of Feynman rules for this Lagrangian, relevant to our calculation, in Appendix~\ref{app:rulesExpr}. 
For the Yang-Mills part they are equivalent to the standard QCD rules and can be taken from the literature (e.g. Ref.~\cite{Romao:2012pq} with $\eta_G=1$, $\eta_s=-1$).
Details of the derivation of the electroweak counterterms and renormalisation are presented in Appendix~\ref{app:renormalisation}.

To evaluate our predictions, we must also specify a consistent electroweak input-parameter scheme.
We take the top-quark mass and Higgs boson mass in the on-shell (OS) scheme as inputs to our calculation.
The renormalised top-quark Yukawa coupling, $g_t$, depends on the top-quark mass and the vev, it is fixed via the relation given in Eq.~\eqref{eq:SMcouplings} after renormalising the top-quark mass and the vev. Similarly, the renormalised trilinear, $g_3$, and quartic, $g_4$, Higgs boson self-couplings depend on the Higgs mass and the vev and are fixed via Eq.~\eqref{eq:SMcouplings} after renormalising the Higgs mass and the vev.
In the gaugeless limit, we can consider the $Z$ and $W$ bosons (which do not appear directly in our computation) to be massless particles; therefore, it is natural to pick $M_Z=0$ and $M_W=0$ as input parameters.
Finally, we must specify the value of the electromagnetic coupling constant.
The most natural choice in our parameterisation would be to specify the value of the vev, $v$, after renormalisation.
However, to simplify the connection to more commonly used input schemes, we instead take $G_F$ as an input parameter and derive from it the value of the vev in the $G_\mu$ (a.k.a. $\alpha_\mu$) scheme. That is to say, we require that the relation $v = (\sqrt{2} G_F)^{-\frac{1}{2}}$ holds to all orders in perturbation theory.
We circumvent the complication that the muon decay vertex employed for the matching in the $G_\mu$ scheme is not present in our model by relying on external calculations from e.g. Ref.~\cite{Biekotter:2023xle} to fix the finite parts of the vev renormalisation; for further details, see Section~\ref{sec:renormalisation} and Appendix~\ref{ssec:dvev}.
In summary, the input-parameter scheme of our calculation is therefore $\left\{ M_Z=0, M_W=0, G_F\right\}$ + $\left\{m_t,m_H\right\}$, where all masses are specified in the on-shell scheme.

\subsection{Amplitude Structure}
\label{sec:amplitude}
We compute the amplitude for the process $g^a_\mu(p_1) g^b_\nu(p_2) \rightarrow H(-p_3) H(-p_4)$, with all momenta defined as incoming.
The amplitude may be parametrised in terms of the usual Mandelstam invariants,
\begin{align}
&s=(p_1 + p_2)^2,& &t=(p_1 + p_3)^2,& &u=(p_2 + p_3)^2,&
\end{align}
with $p_1^2 = p_2^2 = 0$ and $p_3^2 = p_4^2 = m_H^2$.
Due to momentum conservation, $p_1 + p_2 + p_3 + p_4 = 0$, the invariants obey the additional relation $s+t+u = 2 m_H^2$.

As described in Section~\ref{sec:model}, we will consider only the subset of electroweak corrections appearing in the SM, involving the top-quark Yukawa coupling and the Higgs boson trilinear and quartic couplings.
The electroweak $W$ and $Z$ gauge bosons do not appear in our calculation, therefore, no axial-vector couplings are present in the amplitude.
Using QCD gauge invariance, the amplitude may be written in terms of only two independent form factors,
\begin{align}\label{eq:fullamp}
\mathcal{M}_{ab} = 
\varepsilon_{1,\mu} \varepsilon_{2,\nu} M^{\mu \nu}_{ab} =
\varepsilon_{1,\mu} \varepsilon_{2,\nu}
\delta_{ab}
\left(F_1 T_1^{\mu \nu} + F_2 T_2^{\mu \nu} \right),
\end{align}
where $ \varepsilon_{1,\mu},  \varepsilon_{2,\nu}$ are the gluon polarisation vectors, $a,b$ are colour indices in the adjoint representation and $F_1, F_2$ are scalar form factors.
The tensor structures can be chosen to be
\begin{align}
T_1^{\mu \nu} &= g^{\mu \nu} - \frac{p_2^\mu p_1^\nu}{p_1 \cdot p_2}, \\
T_2^{\mu \nu} &= g^{\mu \nu} + \frac{1}{p_T^2 (p_1 \cdot p_2)} \left[ m_H^2 p_2^\mu p_1^\nu - 2 ( p_1 \cdot p_3) p_2^\mu p_3^\nu - 2 (p_2 \cdot p_3) p_3^\mu p_1^\nu + 2 (p_1 \cdot p_2) p_3^\mu p_3^\nu \right],
\end{align}
where $p_T^2 = (u t - m_H^4)/s$ and $T_1 \cdot T_1 = T_2 \cdot T_2 = D-2$, $T_1 \cdot T_2 = D-4$, where $D = 4 - 2 \epsilon$ is the number of space-time dimensions, such that the individual form factors correspond to helicity amplitudes: $\mathcal{M}^{++}=\mathcal{M}^{--}=-F_1$ and $\mathcal{M}^{+-}=\mathcal{M}^{-+}=-F_2$.
The form factors are individually gauge invariant and can be separately renormalised, see Section~\ref{sec:renormalisation}, meaning that the interference contribution between the renormalised form factors vanishes in the limit $\epsilon \rightarrow 0$.

The scalar form factors, $F_i$, can be extracted from the amplitude, $M_{ab}^{\mu \nu}$, using projectors defined to obey the relations,
\begin{align}
\sum_\mathrm{pol} P_{i,a b}^{\mu \nu} \,  \varepsilon^*_{1,\mu} \varepsilon^*_{2,\nu} \varepsilon_{1,\mu^\prime} \varepsilon_{2,\nu^\prime} \delta^{a a^\prime} \delta^{b b^\prime} M^{\mu^\prime \nu^\prime}_{a^\prime b^\prime} = P_{i,a b}^{\mu \nu} M_{\mu \nu}^{a b}=  F_i
\end{align}
with $\sum_\mathrm{pol} \varepsilon^*_{1,\mu} \varepsilon_{1,\mu^\prime} = - g_{\mu \mu^\prime}$.
The projectors are given explicitly by,
\begin{align}
P_{1,ab}^{\mu \nu} &= \frac{\delta_{ab}}{N_c^2-1}\ \frac{1}{4(D-3)} \ \!\left[ (D-2) T_1^{\mu \nu} + (4-D) T_2^{\mu \nu} \right], \label{eq:projector_p1}\\
P_{2,ab}^{\mu \nu} &= \frac{\delta_{ab}}{N_c^2-1}\ \frac{1}{4(D-3)}\  \!\left[ (4-D) T_1^{\mu \nu} + (D-2) T_2^{\mu \nu} \right]. \label{eq:projector_p2}
\end{align}
where the $N_c^2-1$ (with $N_c=3$) appearing in the denominator cancels the colour factor appearing from the $\delta_{ab}$ in the projector contracting with the $\delta_{ab}$ in the decomposed amplitude.

Each of the bare form factors can be expanded in terms of the bare electroweak couplings as follows,
\begin{align} 
F_i &= F_i^{(0)} + F_i^{(1)}, \\
F_i^{(0)} &= g_{s,0}^2  
\Big(
g_{3,0}\, g_{t,0}\ F^{(0)}_{i,g_3 g_t} 
+ g_{t,0}^2\ F^{(0)}_{i,g_t^2} 
\Big), \\
F_i^{(1)} &= g_{s,0}^2  
\Big(
g_{3,0}\, g_{4,0}\, g_{t,0}\  F^{(1)}_{i, g_3 g_4 g_t}
+ g_{3,0}^3\, g_{t,0}\ F^{(1)}_{i, g_3^3 g_t}
+ g_{4,0}\, g_{t,0}^2\ F^{(1)}_{i, g_4 g_t^2} \nonumber \\
& \phantom{= g_s^2 \Big(}
+ g_{3,0}^2\, g_{t,0}^2\ F^{(1)}_{i, g_3^2 g_t^2}
+ g_{3,0}\, g_{t,0}^3\ F^{(1)}_{i, g_3 g_t^3}
+ g_{t,0}^4\ F^{(1)}_{i, g_t^4} 
\Big),\label{eq:couplingStructures}
\end{align}
where $g_s = \sqrt{4 \pi \alpha_s}$ is the strong coupling.
The bare form factors correspond to the coefficients of the bare couplings, we suppress the $0$ subscript of the bare couplings in the labels of the form factors for brevity.
The form factors $F^{(0)}_{i,j}$ correspond to the leading-order one-loop triangle and box contributions, while $F^{(1)}_{i,j}$ correspond to the $6$ possible coupling structure combinations appearing at two loops.
We expand our bare form factors in the electroweak coupling, $\alpha_0 \propto 1/{v_0}^2$, such that the products of couplings entering at LO scale as $1/{v_0}^{2}$ while the products of couplings entering at NLO scale as $1/{v_0}^{4}$.

The bare form factors may be further decomposed into sets of one-particle-irreducible (1PI) and one-particle-reducible (1PR) diagrams.
At leading order, the form factors, split according to combinations of the EW couplings ($g_{t,0}, g_{3,0}, g_{4,0}$), are either entirely 1PI or 1PR,
\begin{align}
    &F_{i,g_3 g_t}^{(0)} = F_{i,g_3 g_t}^{(0),\mathrm{1PR}},&
    &F_{i,g_t^2}^{(0)} = F_{i,g_t^2}^{(0),\mathrm{1PI}}.&
\end{align}
Starting from NLO, the form factors contain a mixture of 1PI and 1PR contributions,
\begin{align}
    &F_{i,g_3^3 g_t}^{(1)} = F_{i,g_3^3 g_t}^{(1),\mathrm{1PR}},&
    &F_{i,g_3 g_4 g_t}^{(1)} = F_{i,g_3 g_4 g_t}^{(1),\mathrm{1PR}},& \\
    &F_{i,g_3^2 g_t^2}^{(1)} = F_{i,g_3^2 g_t^2}^{(1),\mathrm{1PI}} + F_{i,g_3^2 g_t^2}^{(1),\mathrm{1PR}},&
    &F_{i,g_4 g_t^2}^{(1)} = F_{i,g_4 g_t^2}^{(1),\mathrm{1PI}} + F_{i,g_4 g_t^2}^{(1),\mathrm{1PR}},& \\
    &F_{i,g_3 g_t^3}^{(1)} = F_{i,g_3 g_t^3}^{(1),\mathrm{1PI}} + F_{i,g_3 g_t^3}^{(1),\mathrm{1PR}},&
    &F_{i,g_t^4}^{(1)} = F_{i,g_t^4}^{(1),\mathrm{1PI}} + F_{i,g_t^4}^{(1),\mathrm{1PR}}.&
\end{align}  
We compute each of the bare form factors $F^{(0)}_{i,j}$ and $F^{(1)}_{i,j}$ separately and obtain results for both the 1PI and 1PR contributions separately.

At leading order, the partonic cross section can be written as
\begin{equation}
    \hat{\sigma}^{(0)}=\frac{1}{16\pi s^2}\int_{t_-}^{t_+}\mathrm{d}t\left|\overline{\mathcal{M}}^{(0)}\right|^2=\frac{1}{512\pi s^2}\int_{t_-}^{t_+}\mathrm{d}t\left(\left|F_1^{(0)}\right|^2+\left|F_2^{(0)}\right|^2\right)
\end{equation}
where 
\begin{equation}
    t_{\pm}=m_H^2-\frac{s}{2}\left[1\mp\sqrt{1-\frac{4m_H^2}{s}}\ \right]
\end{equation}
are the boundaries in $t$ of the physical region for a given $s\geq4m_H^2$. The averaged matrix element squared $\left|\overline{\mathcal{M}}^{(0)}\right|^2$ contains a symmetry factor for the two final state Higgs bosons, spin and colour averaging for the incoming gluons, a factor of $D - 2$ from the square of the tensor structures ($D-4$ from the interference between the tensor structures does not contribute as explained above) and another factor of $N_c^2-1$ from the sum over the adjoint colour indices of $\delta_{ab}$ in Eq.~\eqref{eq:fullamp}. To obtain the total cross section, the partonic cross section must be convoluted with the parton distribution functions (PDFs) in the usual way.

At NLO in the electroweak expansion, the bare form factors can have UV divergences which give rise to poles of order $1/\epsilon$ which are treated by renormalising the masses and fields of the Higgs boson and top quark along with the vacuum expectation value.
We perform the UV renormalisation by computing explicit counterterm amplitudes, separated on couplings structures, as described in Section~\ref{sec:renormalisation}.
In this way, we retain the complete dependence of our amplitudes on the individual couplings which facilitates changing the electroweak input scheme or supplementing our calculation with higher-dimensional effective field theory operators.
The subset of corrections that we consider here consists of corrections involving the emission of additional massive particles from massive particle lines and is therefore free of IR singularities.

\subsection{Diagram \& Amplitude Generation}
\label{sec:setup}
We generate Feynman diagrams using \textsc{qgraf}~\cite{Nogueira:1991ex} and find a total of 168 diagrams, after excluding tadpole diagrams and diagrams present in the full Standard Model but not in our reduced Yukawa Model.
We generate the amplitude using two independent tool chains based on either a) \textsc{alibrary}~\cite{alibrary}, a \textsc{Mathematica} and \textsc{Form}~\cite{Kuipers:2013pba} package for computing multi-loop amplitudes, or b) \textsc{Reduze 2}~\cite{vonManteuffel:2012np}.
The resulting amplitudes agree up to sector relations and symmetries before applying Integration-By-Parts identities.

\begin{table}[h]
\centering
\begin{tabularx}{1.0\textwidth} { 
  >{\centering\arraybackslash}X
  >{\centering\arraybackslash}X
  >{\centering\arraybackslash}X
  >{\centering\arraybackslash}X 
  >{\centering\arraybackslash}X 
  >{\centering\arraybackslash}X 
  >{\centering\arraybackslash}X }
 \hline
 Type & $g_3g_4g_t$ & $g_3^3g_t$ & $g_4 g_t^2$ & $g_3^2 g_t^2$ & $g_3g_t^3$ & $g_t^4$  \\
 \hline
  1PI & 0  & 0  & 3 & 6  & 24  & 60  \\
  1PR & 12  & 6  & 1 & 6  & 24  & 26  \\
  Total & 12  & 6  & 4 & 12  & 48  & 86  \\
 \hline
\end{tabularx}
\caption{Number of Feynman diagrams (one-particle-irreducible, one-particle-reducible and total), excluding tadpole diagrams, which contribute to each of the bare coupling structures at NLO.}
\label{tab:numdiagrams}
\end{table}

\begin{figure}[ht]
    \centering
    \begin{subfigure}{0.49\textwidth}
    \centering
    \raisebox{0.5ex}{\scalebox{0.75}{\begin{tikzpicture}
	\begin{pgfonlayer}{nodelayer}
		\node [style=dot] (0) at (-2, 1) {};
		\node [style=dot] (1) at (-0.5, 0) {};
		\node [style=dot] (3) at (-2, -1) {};
		\node [style=dot] (4) at (0.7, 0) {};
		\node [style=none] (8) at (-4, -1) {};
		\node [style=none] (9) at (-4, 1) {};
		\node [style=dot] (11) at (2.3, 0) {};
		\node [style=none] (12) at (4, 1) {};
		\node [style=none] (13) at (4, -1) {};
	\end{pgfonlayer}
	\begin{pgfonlayer}{edgelayer}
		\draw [style=fermionarrow] (0) to (1);
		\draw [style=fermionarrow] (3) to (0);
		\draw [style=gluoncoil] (9.center) to (0);
		\draw [style=gluoncoil] (8.center) to (3);
		\draw [style=scalar] (1) to (4);
		\draw [style=fermionarrow] (1) to (3);
		\draw [style=scalar, bend right=90, looseness=1.4] (11) to (4);
		\draw [style=scalar] (13.center) to (11);
		\draw [style=scalar, bend right=90, looseness=1.4] (4) to (11);
		\draw [style=scalar] (11) to (12.center);
	\end{pgfonlayer}
\end{tikzpicture}}}
    \caption{$g_3g_4g_t$}
    \label{sfig:g3g4gt}
    \end{subfigure}
    \hfill
    \begin{subfigure}{0.49\textwidth}
    \centering
    \raisebox{0.5ex}{\scalebox{0.75}{\begin{tikzpicture}
	\begin{pgfonlayer}{nodelayer}
		\node [style=dot] (0) at (-2, 1) {};
		\node [style=dot] (1) at (-0.5, 0) {};
		\node [style=dot] (3) at (-2, -1) {};
		\node [style=dot] (4) at (0.5, 0) {};
		\node [style=none] (8) at (-4, -1) {};
		\node [style=none] (9) at (-4, 1) {};
		\node [style=dot] (10) at (2, 1) {};
		\node [style=dot] (11) at (2, -1) {};
		\node [style=none] (12) at (4, 1) {};
		\node [style=none] (13) at (4, -1) {};
	\end{pgfonlayer}
	\begin{pgfonlayer}{edgelayer}
		\draw [style=fermionarrow] (0) to (1);
		\draw [style=fermionarrow] (3) to (0);
		\draw [style=gluoncoil] (9.center) to (0);
		\draw [style=gluoncoil] (8.center) to (3);
		\draw [style=scalar] (1) to (4);
		\draw [style=fermionarrow] (1) to (3);
		\draw [style=scalar] (4) to (10);
		\draw [style=scalar] (10) to (11);
		\draw [style=scalar] (11) to (4);
		\draw [style=scalar] (13.center) to (11);
		\draw [style=scalar] (10) to (12.center);
	\end{pgfonlayer}
\end{tikzpicture}}}
    \caption{$g_3^3g_t$}
    \label{sfig:g33gt}
    \end{subfigure}\\
    \vspace{0.75cm}
    \begin{subfigure}{0.49\textwidth}
    \centering
    \raisebox{0.5ex}{\scalebox{0.75}{\begin{tikzpicture}
	\begin{pgfonlayer}{nodelayer}
		\node [style=dot] (0) at (-2, 1) {};
		\node [style=dot] (1) at (0, 1) {};
		\node [style=dot] (2) at (0, -1) {};
		\node [style=dot] (3) at (-2, -1) {};
		\node [style=dot] (4) at (2, 0) {};
		\node [style=none] (6) at (4, 1) {};
		\node [style=none] (7) at (4, -1) {};
		\node [style=none] (8) at (-4, -1) {};
		\node [style=none] (9) at (-4, 1) {};
	\end{pgfonlayer}
	\begin{pgfonlayer}{edgelayer}
		\draw [style=fermionarrow] (0) to (1);
		\draw [style=fermionarrow] (1) to (2);
		\draw [style=fermionarrow] (2) to (3);
		\draw [style=fermionarrow] (3) to (0);
		\draw [style=gluoncoil] (9.center) to (0);
		\draw [style=gluoncoil] (8.center) to (3);
		\draw [style=scalar] (1) to (4);
		\draw [style=scalar] (2) to (4);
		\draw [style=scalar] (4) to (6.center);
		\draw [style=scalar] (4) to (7.center);
	\end{pgfonlayer}
\end{tikzpicture}}}
    \caption{$g_4g_t^2$}
    \label{sfig:g4gt2}
    \end{subfigure}
    \hfill
    \begin{subfigure}{0.49\textwidth}
    \centering
    \raisebox{0.5ex}{\scalebox{0.75}{\begin{tikzpicture}
	\begin{pgfonlayer}{nodelayer}
		\node [style=dot] (0) at (-2, 1) {};
		\node [style=dot] (3) at (-2, -1) {};
		\node [style=dot] (4) at (0, 1) {};
		\node [style=none] (6) at (4, 1) {};
		\node [style=none] (7) at (4, -1) {};
		\node [style=none] (8) at (-4, -1) {};
		\node [style=none] (9) at (-4, 1) {};
		\node [style=dot] (10) at (0, -1) {};
		\node [style=dot] (11) at (2, 1) {};
		\node [style=dot] (12) at (2, -1) {};
	\end{pgfonlayer}
	\begin{pgfonlayer}{edgelayer}
		\draw [style=fermionarrow] (3) to (0);
		\draw [style=gluoncoil] (8.center) to (3);
		\draw [style=fermionarrow] (10) to (3);
		\draw [style=fermionarrow] (0) to (4);
		\draw [style=fermionarrow] (4) to (10);
		\draw [style=scalar] (4) to (11);
		\draw [style=scalar] (11) to (6.center);
		\draw [style=scalar] (11) to (12);
		\draw [style=scalar] (7.center) to (12);
		\draw [style=scalar] (12) to (10);
		\draw [style=gluoncoil] (9.center) to (0);
	\end{pgfonlayer}
\end{tikzpicture}}}
    \caption{$g_3^2g_t^2$}
    \label{sfig:g32gt2}
    \end{subfigure}\\
    \vspace{0.75cm}
    \begin{subfigure}{0.49\textwidth}
    \centering
    \raisebox{0.5ex}{\scalebox{0.75}{\begin{tikzpicture}
	\begin{pgfonlayer}{nodelayer}
		\node [style=dot] (0) at (-2, 1) {};
		\node [style=dot] (1) at (-1, 1) {};
		\node [style=dot] (3) at (-2, -1) {};
		\node [style=dot] (4) at (0, 1) {};
		\node [style=none] (6) at (2.75, 1) {};
		\node [style=none] (7) at (2.75, -1) {};
		\node [style=none] (8) at (-4.5, -1) {};
		\node [style=none] (9) at (-4.5, 1) {};
		\node [style=dot] (10) at (0, -1) {};
		\node [style=dot] (11) at (1.75, -1) {};
	\end{pgfonlayer}
	\begin{pgfonlayer}{edgelayer}
		\draw [style=fermionarrow] (0) to (1);
		\draw [style=fermionarrow] (3) to (0);
		\draw [style=gluoncoil] (8.center) to (3);
		\draw [style=gluoncoil] (9.center) to (0);
		\draw [style=fermionarrow] (1) to (4);
		\draw [style=scalar] (4) to (6.center);
		\draw [style=fermionarrow] (10) to (3);
		\draw [style=scalar] (10) to (11);
		\draw [style=scalar] (11) to (7.center);
		\draw [style=scalar] (11) to (1);
		\draw [style=fermionarrow] (4) to (10);
	\end{pgfonlayer}
\end{tikzpicture}}}
    \caption{$g_3g_t^3$}
    \label{sfig:g3gt3}
    \end{subfigure}
    \hfill
    \begin{subfigure}{0.49\textwidth}
    \centering
    \raisebox{0.5ex}{\scalebox{0.75}{\begin{tikzpicture}
	\begin{pgfonlayer}{nodelayer}
		\node [style=dot] (0) at (-2, 1) {};
		\node [style=dot] (3) at (-2, -1) {};
		\node [style=dot] (4) at (0, 1) {};
		\node [style=none] (6) at (4, 1) {};
		\node [style=none] (7) at (4, -1) {};
		\node [style=none] (8) at (-4, -1) {};
		\node [style=none] (9) at (-4, 1) {};
		\node [style=dot] (10) at (0, -1) {};
		\node [style=dot] (11) at (2, 1) {};
		\node [style=dot] (12) at (2, -1) {};
	\end{pgfonlayer}
	\begin{pgfonlayer}{edgelayer}
		\draw [style=fermionarrow] (3) to (0);
		\draw [style=gluoncoil] (8.center) to (3);
		\draw [style=fermionarrow] (10) to (3);
		\draw [style=fermionarrow] (0) to (4);
		\draw [style=scalar] (4) to (10);
		\draw [style=fermionarrow] (4) to (11);
		\draw [style=scalar] (11) to (6.center);
		\draw [style=fermionarrow] (11) to (12);
		\draw [style=scalar] (7.center) to (12);
		\draw [style=fermionarrow] (12) to (10);
		\draw [style=gluoncoil] (9.center) to (0);
	\end{pgfonlayer}
\end{tikzpicture}}}
    \caption{$g_t^4$}
    \label{sfig:gt4}
    \end{subfigure}
    \caption{Example diagrams contributing to each of the $6$ coupling structures on which we separate the bare two-loop amplitude.}
    \label{fig:examplediags}
\end{figure}
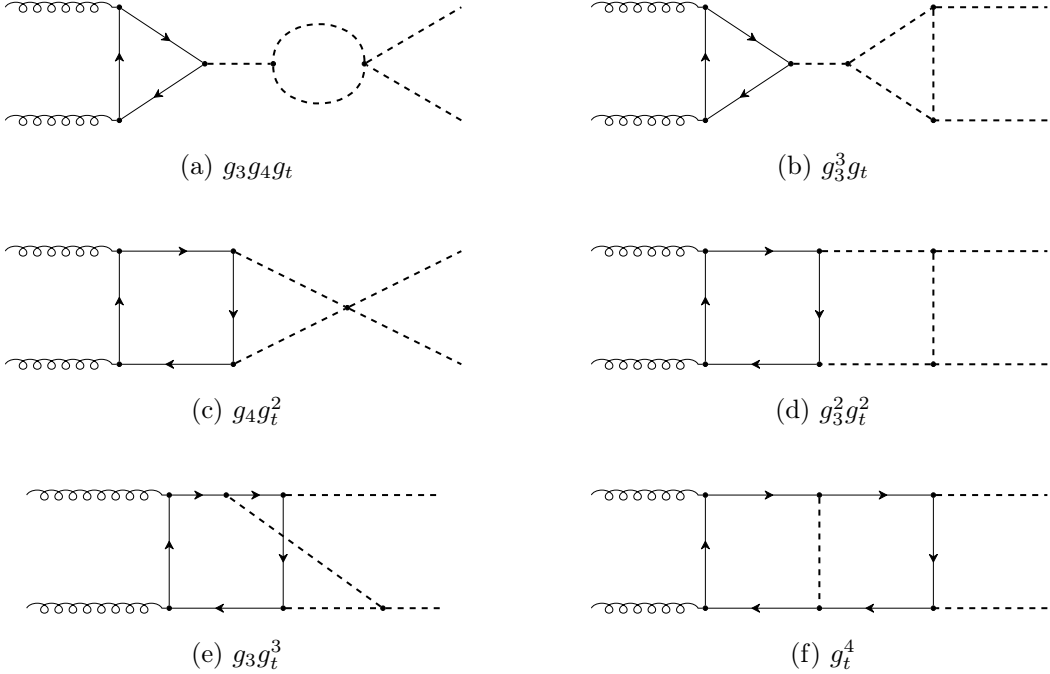

The number of diagrams contributing to each of the coupling structures entering our subset of NLO EW corrections are given in Table~\ref{tab:numdiagrams}, and example Feynman diagrams for each of these structures are shown in Fig.~\ref{fig:examplediags}.
The coupling structures $g_3 g_4 g_t$ and $g_3^3 g_t$ have only a single Yukawa coupling and therefore consist of diagrams that contain loop corrections to the Higgs propagator or trilinear vertex, they are therefore entirely 1PR (see Figs.~\ref{sfig:g3g4gt} and \ref{sfig:g33gt}).
The 1PR contribution to the $g_4 g_t^2$ coupling structure consists of a diagram containing a triple gluon vertex with a single gluon connected to the fermion loop and thus has a vanishing colour factor.
The $g_4 g_t^2$ coupling structure, therefore, receives only a 1PI contribution, see Fig.~\ref{sfig:g4gt2}.
The remaining coupling structures receive contributions from both 1PI and 1PR diagrams.

The complete EW corrections, obtained using the large-$m_t$ limit in Ref.~\cite{Davies:2023npk} and fully using \textsc{AMFlow} in Ref.~\cite{Bi:2023bnq}, contain within them all coupling structures presented in this work, as well as additional contributions from diagrams containing $W$ and $Z$ bosons and their ghosts, as well as the Goldstone bosons.
The coupling structures $g_3 g_4 g_t$ and $g_3^3 g_t$ consist of factorisable one-loop contributions and are comparatively straightforward to compute, they have appeared previously in the literature e.g. Refs.~\cite{Borowka:2018pxx,Bizon:2018syu,Bizon:2024juq}.
The coupling structure $g_4 g_t^2$ contains only three-point integrals, the relevant integrals are known analytically in a large top mass expansion~\cite{Degrassi:2004mx,Degrassi:2016wml}, the complete structure was computed numerically in Ref.~\cite{Borowka:2018pxx}.
The $g_3^2 g_t^2$ coupling structure was also computed numerically in Ref.~\cite{Borowka:2018pxx}.
To the best of our knowledge, the $g_3 g_t^3$ contribution has not been computed separately so far.
The 1PI contribution to $g_t^4$ was computed in the high-energy limit in Ref.~\cite{Davies:2022ram}.

As a cross-check of one of the more challenging pieces of our calculation, we compare the 1PI piece of our bare $g_t^4$ structure to Ref.~\cite{Davies:2022ram}\footnote{We thank the authors of Ref.~\cite{Davies:2022ram} for performing a detailed comparison of our results.} and find good agreement for points at sufficiently high energy $s \gtrsim 4 m_t^2$, with points at $s \gtrsim 9 m_t^2$ differing by less than 2\% and less than 1\% for $s \gtrsim 16 m_t^2$.
The other parts of our calculation are performed systematically using an identical setup to this structure and so are partially checked by this comparison. 
Further checks on our final result are described in Section~\ref{sec:results}.

\subsection{Reduction}
\label{sec:reduction}
\begin{table}
	\centering
	\begin{tabular}{cccc}
		\specialrule{1pt}{0pt}{1pt} \texttt{F1} & \texttt{F2} & \texttt{F3} & \texttt{F4} \\\hline
		$ l_1^2 - m_t^2 $ & $ l_1^2 - m_t^2 $ & $ l_1^2 - m_H^2 $ & $ l_1^2 - m_t^2 $ \\
		$ l_2^2 - m_t^2 $ & $ l_2^2 - m_t^2 $ & $ (l_1 - l_2)^2 - m_t^2 $ & $ l_2^2 - m_H^2 $ \\
		$ (l_1 - l_2)^2 - m_H^2 $ & $ (l_1 - l_2)^2 - m_H^2 $ & $ (l_1 + p_1)^2 - m_H^2 $ & $ (l_1 - l_2)^2 - m_t^2 $ \\
		$ (l_1 + p_1)^2 - m_t^2 $ & $ (l_1 + p_1)^2 - m_t^2 $ & $ (l_2 + p_1)^2 - m_t^2 $ & $ (l_1 + p_1)^2 - m_t^2 $ \\
		$ (l_2 + p_1)^2 - m_t^2 $ & $ (l_2 + p_1)^2 - m_t^2 $ & $ (l_1 - p_2)^2 - m_H^2 $ & $ (l_2 + p_1)^2 - m_H^2 $ \\
		$ (l_1 - p_2)^2 - m_t^2 $ & $ (l_1 - p_3)^2 - m_t^2 $ & $ (l_2 - p_2)^2 - m_t^2 $ & $ (l_1 - p_2)^2 - m_t^2 $ \\
		$ (l_2 - p_2)^2 - m_t^2 $ & $ (l_2 - p_3)^2 - m_t^2 $ & $ (l_2 - p_2 - p_3)^2 - m_t^2 $ & $ (l_2 - p_2)^2 - m_H^2 $ \\
		$ (l_1 - p_2 - p_3)^2 - m_t^2 $ & $ (l_1 - p_2 - p_3)^2 - m_t^2 $ & $ (l_1 + p_1 + p_3)^2 - m_H^2 $ & $ (l_1 - p_2 - p_3)^2 - m_t^2 $ \\
		$ (l_2 - p_2 - p_3)^2 - m_t^2 $ & $ (l_2 - p_2 - p_3)^2 - m_t^2 $ & $ (l_2 + p_1 - p_2)^2 - m_H^2 $ & $ (l_2 - p_2 - p_3)^2 - m_H^2 $\\\specialrule{1pt}{0pt}{1pt}
	\end{tabular}\vspace{5mm}
	\begin{tabular}{ccc}
	\specialrule{1pt}{0pt}{1pt} \texttt{F5} & \texttt{F6} & \texttt{F7} \\\hline
		$ l_1^2 - m_H^2 $ & $ l_1^2 - m_H^2 $ & $ l_1^2 - m_t^2 $ \\
		$ l_2^2 - m_t^2 $ & $ l_2^2 - m_t^2 $ & $ l_2^2 - m_t^2 $ \\
		$ (l_1 - l_2)^2 - m_t^2 $ & $ (l_1 - l_2)^2 - m_t^2 $ & $ (l_1 - l_2)^2 - m_H^2 $ \\
		$ (l_1 + p_1)^2 - m_H^2 $ & $ (l_1 - p_3)^2 - m_H^2 $ & $ (l_1 + p_1)^2 - m_t^2 $ \\
		$ (l_2 + p_1)^2 - m_t^2 $ & $ (l_2 - p_3)^2 - m_H^2 $ & $ (l_2 + p_1)^2 - m_t^2 $ \\
		$ (l_1 - p_3)^2 - m_H^2 $ & $ (l_2 + p_2)^2 - m_t^2 $ & $ (l_1 - p_2)^2 - m_t^2 $ \\
		$ (l_2 - p_3)^2 - m_t^2 $ & $ (l_1 + p_1 + p_2)^2 - m_H^2 $ & $ (l_2 - p_2)^2 - m_t^2 $ \\
		$ (l_1 - p_2 - p_3)^2 - m_H^2 $ & $ (l_1 - l_2 + p_1)^2 - m_t^2 $ & $ (l_1 - l_2 + p_3)^2 - m_H^2 $ \\
		$ (l_2 - p_2 - p_3)^2 - m_t^2 $ & $ (l_1 - l_2 - p_2 - p_3)^2 - m_H^2 $ & $ (l_2 - p_2 - p_3)^2 - m_t^2 $ \\\specialrule{1pt}{0pt}{1pt}
	\end{tabular}
 	\caption[Integral families at NLO]{Integral families used in the reduction (up to permutations of the external legs).}
	\label{tab:intfam}
\end{table}

The loop integrals are written as a list of exponents $\nu_j$ for the denominators $P_j$ of the corresponding integral family $f$ as defined in Table~\ref{tab:intfam}.
In our calculation each loop integral is defined as
\begin{equation}
    \mathcal{I}^{f}_{\vec\nu}(s,t)=\left(\mu^{4-D}\right)^L\int \prod\limits_{i=1}^L\frac{\mathrm{d}^D\ell_i}{i\pi^{D/2}}\prod\limits_{j=1}^N \frac{1}{P_j^{\nu_j}},
\end{equation}
in a general dimension $D$, where $L$ is the number of loops and $N$ is the number of propagators. 
When reporting bare form factors, our integrals are multiplied by an additional factor of $C_{D} =\left(i\pi^{D/2}/(2\pi)^{D}\right)$ per loop, which is required to recover the physical normalisation as dictated by the Feynman rules.

After identifying momentum mapping symmetries with \textsc{Feynson}~\cite{Feynson}, we use a total of 7 integral families (along with permutations of the external legs for 5 of the families) to encode the scalar integrals appearing in all form factors. 
The two-loop families used in this calculation are shown in Table~\ref{tab:intfam}.

To perform the Integration-By-Parts (IBP) reduction~\cite{Chetyrkin:1981qh}, we begin by identifying a suitable basis of master integrals.
Retaining all masses, we find that at two-loop a total of 494 master integrals are required to represent both the NLO amplitude and a closed system of differential equations.
We observe that up to 11 master integrals are required within a single sector, namely, a 6-propagator non-planar sector belonging to family F7 (sector 413 using the ID notation of \textsc{Reduze 2}).

Initially, we choose a finite basis of integrals~\cite{von_Manteuffel_2015} with $D$-factorising denominiators~\cite{Smirnov:2020quc,Usovitsch:2020jrk}, preferring dots over numerators.
Using this basis, the time to numerically evaluate all form factors to a precision of $10^{-3}$ using \textsc{pySecDec} is $\mathcal{O}(100h)$ on a single GPU.
The evaluation time can be decreased by 2-3 orders of magnitude by further optimising the basis choice.
Focusing on the integrals dominating the run time, specifically, the top-level sectors in all integral families and especially those in the most complicated non-planar families (F6 and F7), we searched for a basis in which the masters in the top level ($t=7$) sectors and, where possible, next-to-top level ($t=6$) sectors had coefficients free of poles in the dimensional regulator, $\epsilon$.
To obtain a basis with the required properties we found it necessary to employ both dots and dimensional recurrence relations~\cite{Bern:1993,Tarasov:1996,Lee:2010}.

During the basis search, we found it of practical use to reduce individual sectors neglecting subsectors, thereby avoiding the reconstruction of the vastly more complicated subsector master coefficients, for a large number of different possible basis choices.
With the pole-free coefficient criterion satisfied, we only need to expand the top-level master integrals to leading order in the regulator, vastly reducing the time required to evaluate them numerically.
Furthermore, since we will use the same basis for the evaluation with \textsc{pySecDec} and for the differential equations, we must also avoid poles of the regulator in the ``diagonal'' elements of the differential equation system (as these cannot be removed by similarity transformations of the partial derivative matrices $\mathbf{A}_{x_{i}}$ later).

Our final basis choice consists of integrals with up to three dots expressed in $D$ in the set $\{2-2\epsilon, 4 - 2 \epsilon, 6-2 \epsilon, 8- 2\epsilon\}$.
We could eliminate $1/\epsilon$ poles in the amplitude coefficients for all $t=7$ master integrals and many of the $t=6$ integrals while retaining finiteness and $D$-factorising coefficients for the new basis of integrals\footnote{In our final basis, we have a total of 25 6-propagator master integrals (+25 obtained by crossing) belonging to \texttt{F1}, \texttt{F2} and \texttt{F4} with a $1/\epsilon$ present in their coefficient in the amplitude. 
The integrals $\mathcal I^{\texttt{F5},D=6-2\epsilon}_{(0,1,1,1,1,0,2,1,0)}(s,t)$ and $\mathcal I^{\texttt{F5},D=6-2\epsilon}_{(0,1,1,1,1,0,2,1,0)}(s,u)$ are also present in our final basis, though they are neither finite nor quasi-finite, starting at order $1/\epsilon$. We find that these integrals do not contribute significantly to the evaluation time and did not attempt to improve our basis of master integrals further. However, this would be possible in principle.}.
Crucially, to obtain a basis with these properties, we found it necessary to select integrals in different numbers of dimensions within a single sector.

Having settled on an improved basis, we generate the dimensional recurrence relations and differential equations of the master integrals using \textsc{Reduze 2}, firstly with all in $D_0=\nolinebreak4-2\epsilon$. 
We generate IBP equations with \textsc{Kira}~\cite{Maierhofer:2017gsa,Klappert:2020nbg} covering all integrals appearing in the amplitude, differential equations and dimensional recurrence relations, again in $D_0$. 
Next, we replicate these equations with the relevant shifts of $D_0$ (that is to say, $\pm2n$) to cover the entire system, such that we have enough information to express integrals in any of our equations in terms of masters in any of the relevant dimensions. 
We can load this entire set of equations along with the unreduced amplitude split on coupling structures into \textsc{Ratracer}~\cite{Magerya:2022hvj} and, defining our choice of masters, we can solve this system of equations using \textsc{Kira}, \textsc{Ratracer}, and \textsc{Firefly}~\cite{Klappert:2019emp,Klappert:2020aqs} to express our amplitude and differential equations in terms of our preferred basis of integrals.

We stress that the reduction is obtained fully symbolically, retaining all masses and invariants ($m_t$ is set to 1 in our reduction, but can be restored by dimensional analysis).
Using the same setup, we also obtain a reduction with $m_H^2/m_t^2 = 12/23$.
We find the total size of the rational coefficients in the reduced amplitude to be 99Gb for the fully symbolic reduction and 8.5Gb with the numeric mass ratio inserted, when separated on coupling structures as in Eq.~\eqref{eq:couplingStructures}.

As a cross-check of the reduction and our amplitude, we independently perform a reduction to a different set of masters with the Higgs boson mass set to a numerical constant and confirm the value of our amplitude after the numerical evaluation of the master integrals.
We further checked the integral reduction by obtaining reductions for individual phase-space points, by substituting all kinematic invariants and masses with randomly selected rational values.

\subsection{Evaluation of the Master Integrals}
\label{sec:integrals}
To evaluate the master integrals appearing in our two-loop amplitudes, we rely on the method of sector decomposition, as implemented in the latest version of \textsc{pySecDec}.
We first generate expressions for the reduced amplitudes in terms of the 494 master integrals, as described in Section~\ref{sec:reduction}.
The amplitudes along with the definitions of the integral families are passed to \textsc{pySecDec}, which generates a single code capable of evaluating all bare form factors $F^{(1)}_{i,j}$, with $i=1,2$ and $j=\{g_3 g_4 g_t,\,g_3^2 g_t,\, g_4 g_t^2,\, g_3^2 g_t^2,\, g_3 g_t^3,\, g_t^4\}$.
The code is automatically generated such that the master integrals are numerically evaluated only once per phase-space point and then used to generate results for each of the form factors and coupling structures.

Our amplitude code is generated by retaining the full symbolic dependence on $s$, $t$, $m_t$, $m_H$ and expanding in $\epsilon$.
When evaluating phase-space points, in order to obtain numerically stable coefficients, it is necessary to insert the Mandelstam invariants and masses in a precision higher than the usual floating point double precision.
In our code, the input values for the Mandelstam invariants and masses are cast to rational numbers by picking the smallest fraction which reproduces $s/m_t^2$ and $t/m_t^2$ to a precision of $10^{-5}$, we also set $m_H^2 = 12/23$, with $m_t^2 = 1$.
We stress, however, that since we have retained the full symbolic dependence on the masses in the integral reduction and the generation of our code, we can therefore arbitrarily vary the value of the Higgs boson and top-quark masses.

Due to the significant size of the rational coefficients present in our fully symbolic amplitude, the evaluation of the master integral coefficients can itself be time-consuming, taking a few minutes to obtain the numeric value of all of the master integral coefficients.
We, therefore, find it beneficial to generate a second code with the specific value for the Higgs boson mass pre-inserted into the coefficients, this reduces the time taken to evaluate the master integral coefficients significantly.

Upon integration, with the master integral basis we have chosen, we observe spurious poles up to order $\epsilon^{-4}$ in the coupling structures $g_3^2g_t^2$, $g_3g_t^3$ and $g_t^4$.
Upon integration, the coefficient of the $\epsilon^{-4}, \epsilon^{-3}$ and $\epsilon^{-2}$ poles vanish within the precision of the numerical integration, leaving a non-zero $\epsilon^{-1}$ pole (for structures $\{g_3g_4g_t,g_3^3g_t,g_3g_t^3,g_t^4\}$ in form factor $F_1^{(1)}$ and for $g_t^4$ in $F_2^{(1)}$) and finite part.
The remaining UV $\epsilon^{-1}$ pole is cancelled against the corresponding counter-term amplitude only after integration.

When evaluating the amplitude, \pysecdec\  adaptively adjusts the precision with which each integral is obtained in order to reach a given precision for the amplitude (more specifically, each form factor, $F_{i,j}^{(k)}$) in the minimum time.
This means that complicated (slow to evaluate or slow to converge) integrals are typically sampled less by the algorithm unless they dominate the uncertainty estimate on the amplitude.
In contrast, the algorithm may spend more time evaluating simple integrals precisely, if their contribution to the amplitude is large.
In our production runs, we request a relative precision of $10^{-3}$ on the finite part of each two-loop form factor for each coupling structure, $F_{i,j}^{(1)}$.

For a typical bulk phase-space point, with $s\approx 561/130 \cdot m_t^2$ and $t\approx-566/217 \cdot m_t^2$, the integration takes approximately five minutes on four GPUs\footnote{Nvidia A100-PCIE-40GB, CUDA v12040}.
For the selected phase-space point, the algorithm spends the most time evaluating the integrals $\mathcal I^\texttt{F4}_{(1,0,1,1,1,1,1,0,1)}(s,t)$ and $\mathcal I^\texttt{F4}_{(1,0,1,1,1,2,0,0,1)}(s,t)$ and uses the most integrand evaluations for $\mathcal I^\texttt{F4}_{(1,0,1,1,1,1,1,0,1)}(s,t)$ and $\mathcal I^\texttt{F4}_{(1,0,1,1,1,1,1,0,1)}(s,u)$.
The least precisely known integrals are $\mathcal I^\texttt{F4}_{(1,0,1,1,1,1,1,0,0)}(s,t)$ with an uncertainty of $\mathcal{O}(3\times10^{-4})$ and $\mathcal I^\texttt{F1}_{(1,0,1,1,1,1,1,0,0)}(s,t) $ with an uncertainty of $1\times10^{-4}$, followed by $\mathcal I^\texttt{F4}_{(1,0,0,1,2,1,1,0,0)}(s,t)$ with an uncertainty around $6\times10^{-5}$. 
For a point in the high energy regime with $s\approx123\cdot m_t^2$ and $t\approx-7/5\cdot m_t^2$ we find that the integrals $I^\texttt{F4}_{(1,0,1,0,2,1,0,1,1)}(s,t)$ and $I^\texttt{F4}_{(1,0,1,0,2,1,0,1,1)}(s,t)$ are the least precisely known after an integration time of two hours. Up to this point, the most time was spent on the integrals $\mathcal I^\texttt{F4}_{(1,0,1,1,1,2,0,0,1)}(s,u)$ and $\mathcal I^\texttt{F1}_{(0,1,1,1,1,0,2,1,0)}(s,u)$, which are also sampled the most. 
We remark that all of these integrals are planar.

\begin{figure}
    \centering
    \includegraphics[width=0.6\textwidth]{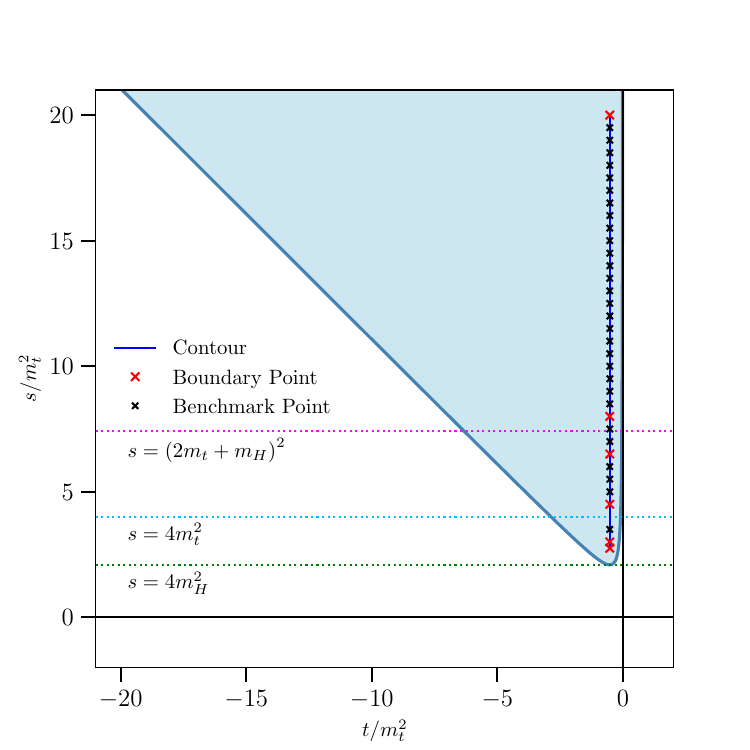}
    \caption{The physical region in the $s$-$t$ plane with the physical thresholds corresponding to $s$-channel cuts shown with dotted lines. Our test contour increasing in $s$ is shown in blue with boundary points plotted along with benchmark points verified in \pysecdec{}.}
    \label{fig:contour}
\end{figure}

\begin{figure}
\centering
    \begin{subfigure}{0.495\textwidth}
    \centering
    \includegraphics[width=1.12\textwidth]{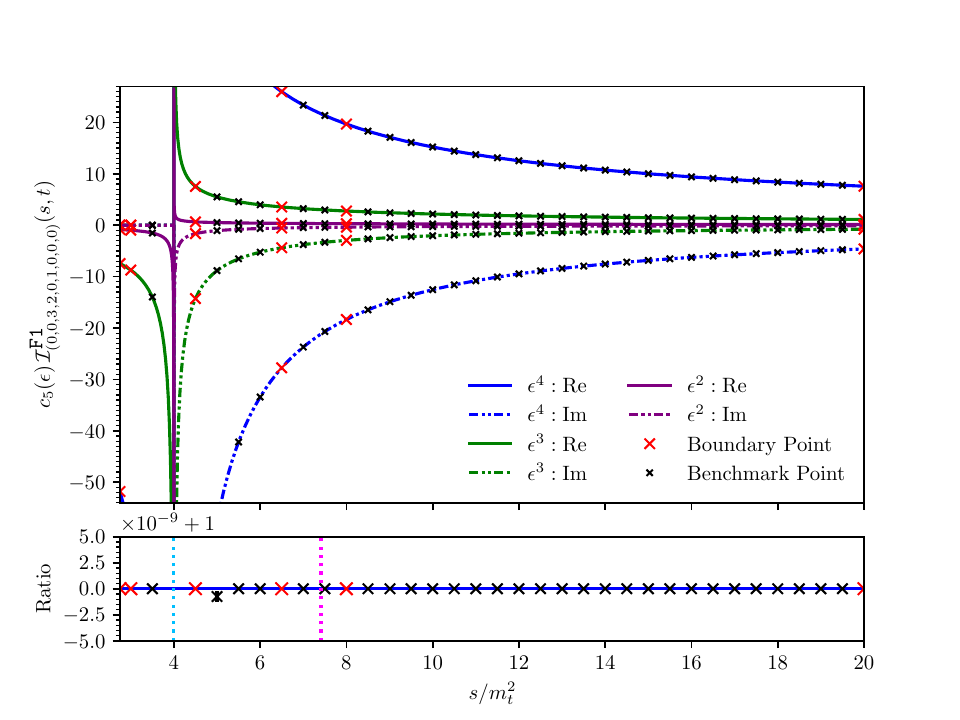}
    \caption{An integral which diverges at the $t \overline{t}$ threshold.}
    \label{sfig:coeffa}
    \end{subfigure}
    \hfill
    \begin{subfigure}{0.495\textwidth}
    \centering
    \includegraphics[width=1.12\textwidth]{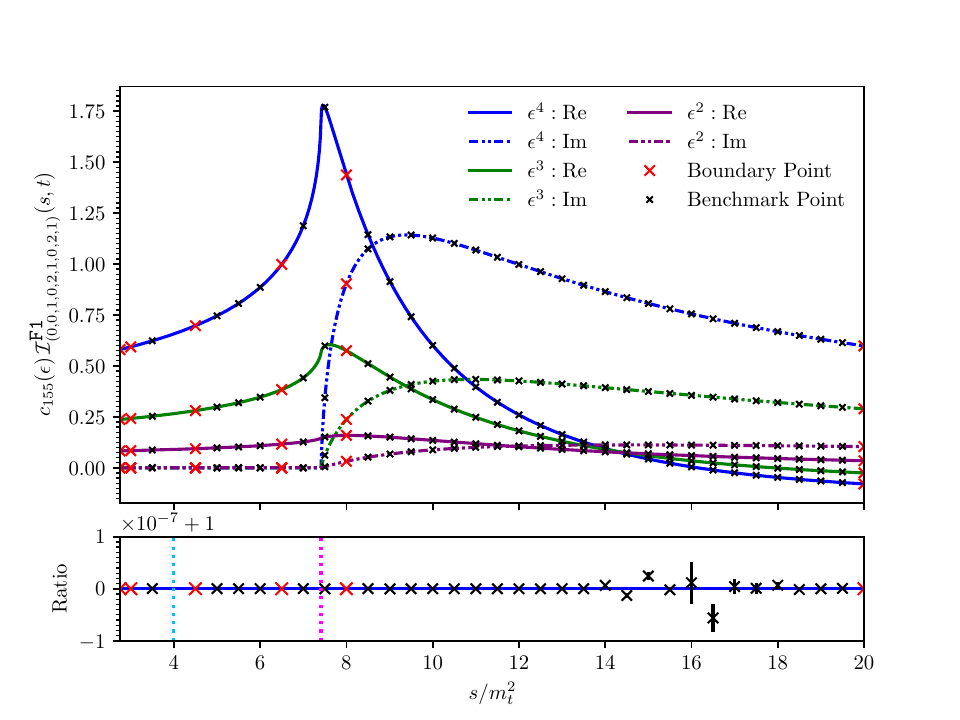}
    \caption{An integral with a $t \overline{t} H$ threshold.}
    \label{sfig:coeffb}
    \end{subfigure}\\
    \begin{subfigure}{0.495\textwidth}
    \centering
    \includegraphics[width=1.12\textwidth]{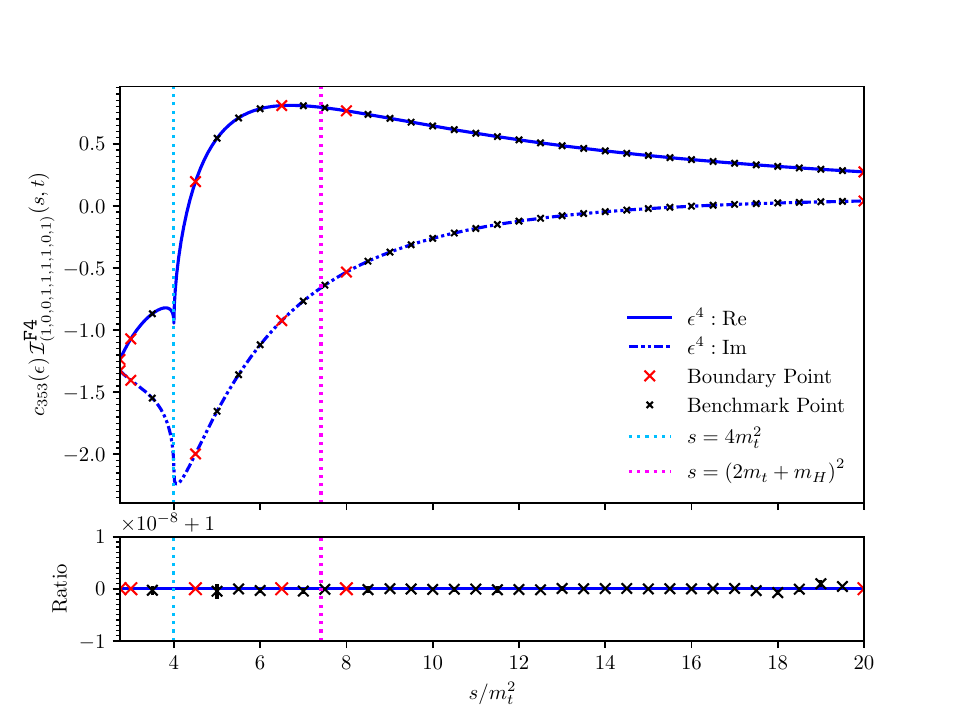}
    \caption{An integral with a $t\overline{t}$ threshold.}
    \label{sfig:coeffc}
    \end{subfigure}
    \hfill
    \begin{subfigure}{0.495\textwidth}
    \centering
    \includegraphics[width=1.12\textwidth]{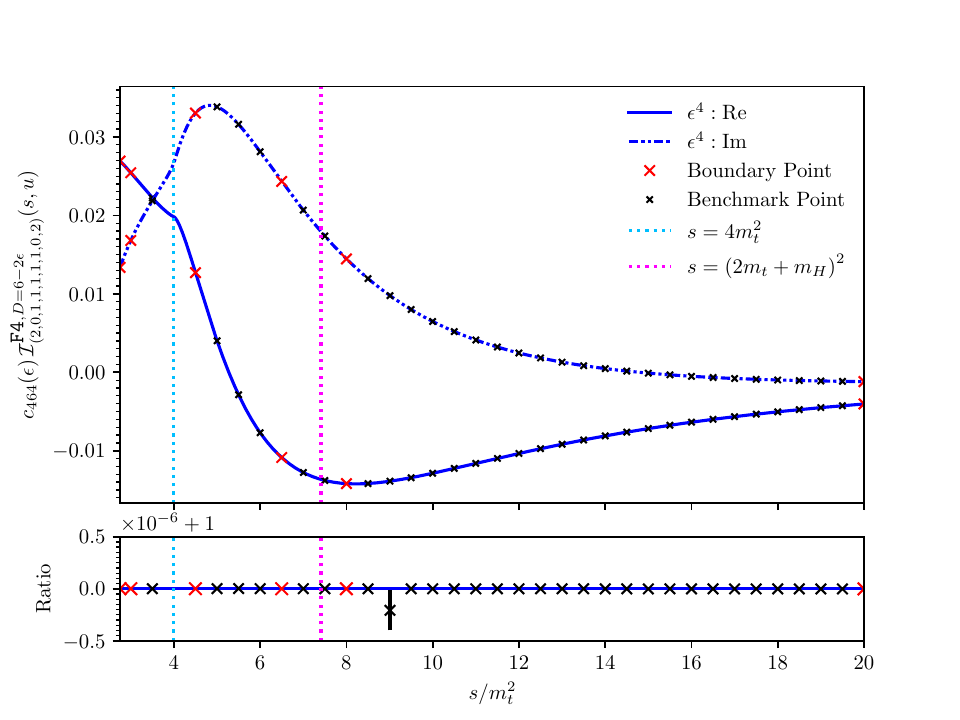}
    \caption{A $t=7$ top-level integral.}
    \label{sfig:coeffd}
    \end{subfigure}
\caption{Real and imaginary parts of coefficients in the $\epsilon$-expansion of selected rescaled master integrals taken along the contour shown in Fig.~\ref{fig:contour}. \subref{sfig:coeffa}) Rescaled master \#5: $c_5\!\left(\epsilon\right)\mathcal I^\texttt{F1}_{(0,0,3,2,0,1,0,0,0)}(s,t)$, \subref{sfig:coeffb}) rescaled master \#155: $c_{155}\!\left(\epsilon\right)\mathcal I^\texttt{F1}_{(0,0,1,0,2,1,0,2,1)}(s,t)$, \subref{sfig:coeffc}) rescaled master \#353: $c_{353}\!\left(\epsilon\right)\mathcal I^\texttt{F4}_{(1,0,0,1,1,1,1,0,1)}(s,t)$ and \subref{sfig:coeffd}) rescaled master \#464: $c_{464}\!\left(\epsilon\right)\mathcal I^{\texttt{F4},D=6-2\epsilon}_{(2,0,1,1,1,1,1,0,2)}(s,u)$. The lower panel of each figure shows the ratio of the {\sc pySecDec} result to the {\sc DiffExp} result for the real part of the coefficient of $\epsilon^4$ which contributes to the amplitude at finite order.}
\label{fig:DiffExpResults}
\end{figure}

As a cross-check of the numerical evaluation of our master integrals, we have also obtained a set of differential equations which are symbolic in $s$ and $t$ and have the aforementioned numeric values for the masses. The differential equations are obtained for the same master integral basis as selected for the numerical evaluation described above and are therefore not in canonical form. They are then rescaled by rational functions of $\epsilon$ to eliminate poles in $\epsilon$ in the differential equation matrices. To verify our numerical evaluation, we use \pysecdec{} to generate a number of boundary points at high precision along a contour of increasing $s$ in the $s$-$t$ plane as in Fig.~\ref{fig:contour}. We then run smaller contours in {\sc DiffExp} between these boundary points to obtain results for the entire contour (except for the $t\overline{t}H$ threshold where we run very close to the threshold above and below without ever crossing it). We check at given benchmark points that the evaluations from {\sc DiffExp} and \textsc{pySecDec} are consistent and plots of a selection of rescaled master integrals are shown in Fig.~\ref{fig:DiffExpResults}. The real and imaginary parts of the coefficients of the required orders of $\epsilon$ in the expansion of the rescaled master integrals are plotted along with the corresponding boundary and benchmark points. A ratio of the \pysecdec{} result to the {\sc DiffExp} result is given in the lower subplot. For completeness, we list the rescalings of the selected master integrals here:
\begin{align*}
&c_{5}\!\left(\epsilon\right)=\frac{\epsilon }{(\epsilon -1)^2 (2 \epsilon -3) (2 \epsilon
   -1)^2 (2 \epsilon +1) (3 \epsilon -2) (3 \epsilon -1) (4
   \epsilon -3) (4 \epsilon -1)},\\
    &c_{155}\!\left(\epsilon\right)=\frac{\epsilon ^2}{2 \epsilon -1},\qquad\quad\quad
    c_{353}\!\left(\epsilon\right)=\frac{\epsilon ^4}{2 \epsilon -1},\qquad\quad\quad
    c_{464}\!\left(\epsilon\right)=\frac{\epsilon ^4}{2 \epsilon -1}.
\end{align*}
Details of the analytic continuation procedure for the master integrals are given in Appendix~\ref{app:continuation}.

\subsection{Electroweak Renormalisation}
\label{sec:renormalisation}
At higher orders in electroweak theory, a tadpole renormalisation has to be performed on top of the usual field, mass and vertex renormalisation. Since the gaugeless limit removes the coupling $\alpha$ from the theory, conventional input parameter schemes that involve $\alpha$ cannot be used. 
As described in Section~\ref{sec:model}, we fix the input parameters $m_H$ and $m_t$ in the on-shell scheme and use the $G_\mu$ scheme for the vev.
Tadpole contributions are treated within the Fleischer-Jegerlehner tadpole scheme (FJTS)~\cite{Fleischer:1980ub}.

The $ G_\mu $ scheme imposes the renormalisation condition that the expression for muon decay corresponds at all orders to the effective four-fermion tree-level interaction in Fermi's theory, thereby fixing the relation between $ G_\mu $ and the renormalised vev, and determining the relation between the bare, $v_0$, and renormalised vev, $v$, at each order. 
In the Yukawa model utilised here, the vertex required for muon decay is not present, therefore it is not possible to directly derive the $G_\mu$ scheme relation between the bare and renormalised vev.
In principle, the renormalisation constant for the vev can be fixed from any electroweak vertex in the theory, for example, the triple and quartic Higgs self-interaction vertices or the Yukawa vertex, by requiring that the higher order electroweak corrections to the vertices are finite.
The consistency of the electroweak theory means that the pole part of the vev renormalisation constant matches in all schemes and independently of which vertex is used to fix it, only the finite part differs.
However, to facilitate the use of our result and its interpretation, we present our main results using the $ G_\mu $ scheme, we obtain the finite parts of the vev counter term using the complete expression for the vev counterterm presented in Ref.~\cite{Biekotter:2023xle} in the limit $M_W \rightarrow 0, M_Z \rightarrow 0$.
In Appendix~\ref{ssec:dvev} we derive the vev renormalisation constants from each of the vertices in our theory and discuss this point in further detail.
With the chosen schemes and conditions, the renormalised quantities and counterterms
\begin{align}
	H_0 &= \sqrt{Z_H}H=\sqrt{1+\delta_H}H\label{eq:H0},\\
	t_0 &= \sqrt{Z_t}t=\sqrt{1+\delta_t}t,\\
	m_{H,0}^2 &= m_H^2(1+\delta m_H^2),\\
	m_{t,0} &= m_t(1+\delta m_t)\label{eq:mt0},\\
	v_0 + \Delta v &= v(1+\delta_v) + \Delta v\label{eq:vev0},
\end{align}
can be fixed. Note that the vev renormalisation condition contains one contribution, $ \Delta v $, from the FJTS for the shift of the vev and another contribution, $ \delta_v $, from the vertex correction. For a detailed description of the procedure, please refer to Appendix~\ref{app:renormalisation}. The explicit expressions used in this work are listed in Appendix~\ref{app:rulesExpr} and a comparison to expressions in the literature is presented in Appendix~\ref{app:comparison}.

A gluon field renormalisation factor, $ Z_g $, or strong coupling renormalisation is not needed, since it would only receive electroweak corrections at $ \mathcal{O}(\alpha_s\alpha) $, \textit{i.e.} at a higher order in the strong interaction as a mixed correction, which we do not consider here.

The renormalised amplitude $\mathcal{M}_\mathrm{ren}$ can now be calculated from the sum of the LO matrix element $\mathcal{M}^{(0)}$, with the bare fields and parameters expressed in terms of the renormalised quantities, and the NLO matrix element $\mathcal{M}^{(1)}$,
\begin{equation}
\mathcal{M}_\mathrm{ren} = Z_H\cdot\left[\mathcal{M}^{(0)}\left(m_t\left(1 + \delta m_t\right), m_H^2\left(1 + \delta m_H^2\right), v\left(1 + \delta_v\right) + \Delta v\right) + \mathcal{M}^{(1)}\left(m_t, m_H^2, v\right) \right]~,\label{eq:Mren}
\end{equation}
where we have suppressed the gluon colour indices of the matrix element appearing in Eq.~\eqref{eq:fullamp}.
The matrix element $\mathcal{M}^{(0)}$ contains all one-loop contributions as well as diagrams with counterterm insertions. All occurring parameters are the renormalised ones. Expanding to first order in $\delta X$ with $X=\{H,t,m_H^2, m_t, v\}$ and including the tadpole corrections to the vev, $\Delta v$, we may rewrite the renormalised amplitude as,
\begin{align}
\mathcal{M} _\mathrm{ren} = 
\mathcal{M}^{(0)}(m_t, m_H^2, v)
+ \mathcal{M}^{(1)}_\mathrm{\delta X}(m_t, m_H^2, v) 
+ \mathcal{M}^{(1)}(m_t, m_H^2, v) + \mathcal{O}(\delta X^2),
\end{align}
with
\begin{align}
\mathcal{M}^{(1)}_\mathrm{\delta X} = & \delta_H \mathcal{M}^{(0)}(m_t, m_H^2, v)
+ \delta {m_t}  \mathcal{M}_{\delta m_t}^{(0)}(m_t, m_H^2, v)
+ \delta {m_H^2}  \mathcal{M}_{\delta m_H^2}^{(0)}(m_t, m_H^2, v) \nonumber \\
& + \delta {v}  \mathcal{M}_{\delta v}^{(0)}(m_t, m_H^2, v)
+ \Delta v \mathcal{M}_{\Delta v}^{(0)}(m_t, m_H^2, v). \label{eq:counterterm_amplitude}
\end{align}
In practice, we separate the counterterm amplitudes, $\mathcal{M}^{(1)}_\mathrm{\delta X}$, according to the form factor, $i=1,2$, and the coupling structure, $j=g_3 g_t, g_t^2$,  appearing in the amplitude as well as the additional coupling structures appearing in the counterterms $\delta X$ themselves.
We obtain the finite one-loop and two-loop renormalised form factors by taking the combination,
\begin{align}
F_i^{(0),\mathrm{fin}} &= F_i^{(0)}, \\
F_i^{(1),\mathrm{fin}} &= F_i^{(1)} + F_i^{(1),\delta X},
\end{align}
respectively, where $F_i^{(1),\delta X}$ collects the counterterm contribution obtained by applying the projectors (given in Eqs.~\eqref{eq:projector_p1} and \eqref{eq:projector_p2}) to the counterterm amplitude in Eq.~\eqref{eq:counterterm_amplitude}.
The counterterm amplitudes are generated by inserting the counterterm vertices given in Appendix~\ref{app:rulesExpr} into the one-loop amplitude, leaving the counterterms $\delta X$ and $\Delta v$ symbolic, then factoring them out of the amplitude.
The counterterms $\delta X$ and the tadpole terms $\Delta v$ contain $1/\epsilon$ divergences, therefore, the counterterm amplitudes must be expanded up to and including $\mathcal{O}(\epsilon)$ in order to obtain correct results for $\mathcal{M}^{(1)}_\mathrm{\delta X}$ at finite order.
We evaluate the counterterm amplitudes numerically in \pysecdec, and insert the $A_0$ and $B_0$ (tadpole and bubble) integrals appearing in the counterterms symbolically.

We remark that, when considering the form factors separated by individual coupling structures, the sum of the two-loop and corresponding counter term contribution is not finite for all form factors, i.e. the sums $F_{i,j}^{(1)} + F_{i,j}^{(1),\delta X}$
are not individually finite.
The reason for this is that, in the SM, the couplings $g_t, g_3$, and $g_4$ are not independent quantities, they must obey the relations given in Eq.~\eqref{eq:SMcouplings} for the $\epsilon$ poles of the renormalised amplitude to cancel.
As a result, the Higgs boson trilinear, $g_3$, and quartic, $g_4$, couplings can not be naively varied without also modifying the underlying theory, for example by adding mass dimension-6 (and/or dimension-8) operators, see e.g. Ref.~\cite{Maltoni:2018ttu}, or additional particles/symmetries.

A further complication arises when considering the renormalised form factors separated by individual coupling structures in $G_\mu$ scheme.
As described above, in this scheme the $\delta_v$ counterterm is derived using the muon decay process, which requires the presence of $W$ and $Z$ bosons and their associated couplings in the theory.
Using results from the literature for this counter term, it is not straightforward to separate the contributions to $\delta_v$ in the $G_\mu$ scheme according to the couplings $g_3, g_4$ and $g_t$. 
The contributions of different coupling structures to the $\delta_v$ counterterm also depend on the choice of the renormalisation conditions, as described in Appendix~\ref{ssec:dvev}.
In this work, we therefore do not attempt this separation and report only the finite renormalised form factors $F_i^{(1),\mathrm{fin}}$ constructed employing the SM values/relations for the various couplings.

\section{Results}
\label{sec:results}
In this section, we present the results of our computation.
We begin by discussing both the bare, $F_1$ and $F_2$, and renormalised, $F_1^\mathrm{fin}$ and $F_2^\mathrm{fin}$, form factors, before presenting results for the total cross section and differential distributions at NLO$^\mathrm{EW}$, including only the Yukawa and self-coupling contributions.

\begin{table}
    \centering
    \begin{tabularx}{\textwidth}{|D|C|C|}
        \hline
        \makecell{$j$: $\epsilon^{\#}$}   &\makecell{$F^{\left(1\right)}_{1,j}$}     &\makecell{$F^{\left(1\right)}_{2,j}$}     \\[0.8mm] \hline
        \makecell{$g_3g_4g_t:\epsilon^{-1}$}         &\makecell{$+7.317018424938384\cdot10^{-5}$\\$+3.530994674708006\cdot10^{-5}\,i$}      &\makecell{$0$}         \\ \hline
        \makecell{$g_3g_4g_t:\epsilon^{0}$}         &\makecell{$+3.273276184619130\cdot10^{-4}$\\$+2.941949902790170\cdot10^{-4}\,i$}      &\makecell{$0$}         \\ \hline
        \makecell{$g_3^3g_t:\epsilon^{-1}$}         &\makecell{$+4.035301063033099\cdot10^{-6}$\\$+1.947326866890242\cdot10^{-6}\,i$}      &\makecell{$0$}         \\ \hline
        \makecell{$g_3^3g_t:\epsilon^{0}$}         &\makecell{$+3.4949862900129389\mathbf{51}\cdot10^{-5}$\\$-4.4770066132017743\mathbf{40}\cdot10^{-5}\,i$}      &\makecell{$0$}         \\ \hline
        \makecell{$g_4g_t^{2}:\epsilon^{0}$}         &\makecell{$+1.47015556537543\mathbf{24}\cdot10^{-4}$\\$-3.14685463406167\mathbf{29}\cdot10^{-4}\,i$}      &\makecell{$0$}         \\ \hline
        \makecell{$g_3^2g_t^2:\epsilon^{0}$}         &\makecell{$-3.00418959847\mathbf{12}\cdot10^{-4}$\\$+1.36208618462\mathbf{96}\cdot10^{-4}\,i$}      &\makecell{$-1.0678083\mathbf{12}\cdot10^{-6}$\\$+4.8255108\mathbf{99}\cdot10^{-6}\,i$}         \\ \hline
        \makecell{$g_3g_t^3:\epsilon^{-1}$}         &\makecell{$+9.6208688168\mathbf{78}\cdot10^{-5}$\\$-1.1571837975\mathbf{79}\cdot10^{-4}\,i$}      &\makecell{$0$}         \\ \hline
        \makecell{$g_3g_t^3:\epsilon^{0}$}         &\makecell{$+7.723391320\mathbf{21}\cdot10^{-4}$\\$+1.229726636\mathbf{23}\cdot10^{-4}\,i$}      &\makecell{$+5.947229\mathbf{62}\cdot10^{-5}$\\$+6.546467\mathbf{67}\cdot10^{-5}\,i$}         \\ \hline
      \makecell{$g_t^4:\epsilon^{-1}$}         &\makecell{$-4.5097091352236\mathbf{40}\cdot10^{-3}$\\$-1.0090262890534\mathbf{41}\cdot10^{-3}\,i$}      &\makecell{$-5.411414111\mathbf{26}\cdot10^{-5}$\\$+7.833751223\mathbf{26}\cdot10^{-5}\,i$}         \\ \hline
      \makecell{$g_t^4:\epsilon^{0}$}         &\makecell{$-2.1195755326\mathbf{56}\cdot10^{-2}$\\$-8.8277696639\mathbf{82}\cdot10^{-3}\,i$}      &\makecell{$-3.36569\mathbf{13}\cdot10^{-4}$\\$+4.63388\mathbf{99}\cdot10^{-4}\,i$}         \\ \hline
        
    \end{tabularx}
    \caption{Numeric results for the bare form factors, $F_{i,j}^{(1)}$, for each coupling structure on the phase-space point: \{$s=\nolinebreak799/125$, $t=\nolinebreak-519/500$, $m_H^{2}=\nolinebreak12/23$, $m_t^{2}=\nolinebreak1$\}. Boldface digits represent the error on the final two stated digits and where there are none, the stated digits are accurate to the given precision. Missing $\epsilon$ orders are understood to be identically zero.}
\label{tab:psp_results}
\end{table}

In Table~\ref{tab:psp_results}, we provide explicit numbers for the NLO$^\mathrm{EW}$ contributions of each of the coupling structures to the bare amplitude form factors $F_1$ and $F_2$. We note that these numbers are the coefficients of the coupling structures (and so need to be multiplied by the coupling structures themselves as in Eq.~\eqref{eq:couplingStructures}).
We make a number of comments about these results.
Firstly, we note that $F_1$ and $F_2$ correspond to the $\mathcal{M}^{++}$ and $\mathcal{M}^{+-}$ helicity amplitudes, respectively.
Therefore, contributions to $F_1$ have an initial state with a total spin of zero, whilst $F_2$ receives contributions from initial states with total spin two.
For the structures $g_3 g_4 g_t$ and $g_3^3 g_t$, the contributions to $F_2^{\left(1\right)}$ are zero because these diagrams are entirely one-particle-reducible (1PR) via a cut through a Higgs boson propagator (see Table~\ref{tab:numdiagrams} and Fig.~\ref{fig:examplediags}), therefore the initial states have total spin zero. 
Similarly, the contribution from structure $g_4 g_t^2$ to $F_2^{(1)}$ is zero since diagrams with this structure can only contribute to spin zero due to their symmetry. 
Finally, there is no $1/\epsilon$ pole contribution to $F_2^{\left(1\right)}$ from structure $g_3 g_t^3$ because the only 1PI counterterm diagrams (which must topologically be LO box diagrams to contribute to $F_2^{(1)}$) which could correspond to this coupling structure have counterterm insertions in the Yukawa vertex, the relevant part of the correction is given in Fig.~\ref{fig:vevsubfig} in Appendix~\ref{ssec:dvev}. 
This particular contribution to the Yukawa vertex correction is $\epsilon$-finite, hence this structure's contribution to $F_2^{\left(1\right)}$ is also finite.

\begin{figure}[h!]
\centering
\begin{subfigure}{0.49\textwidth}
\centering\includegraphics[width=1.0\textwidth]{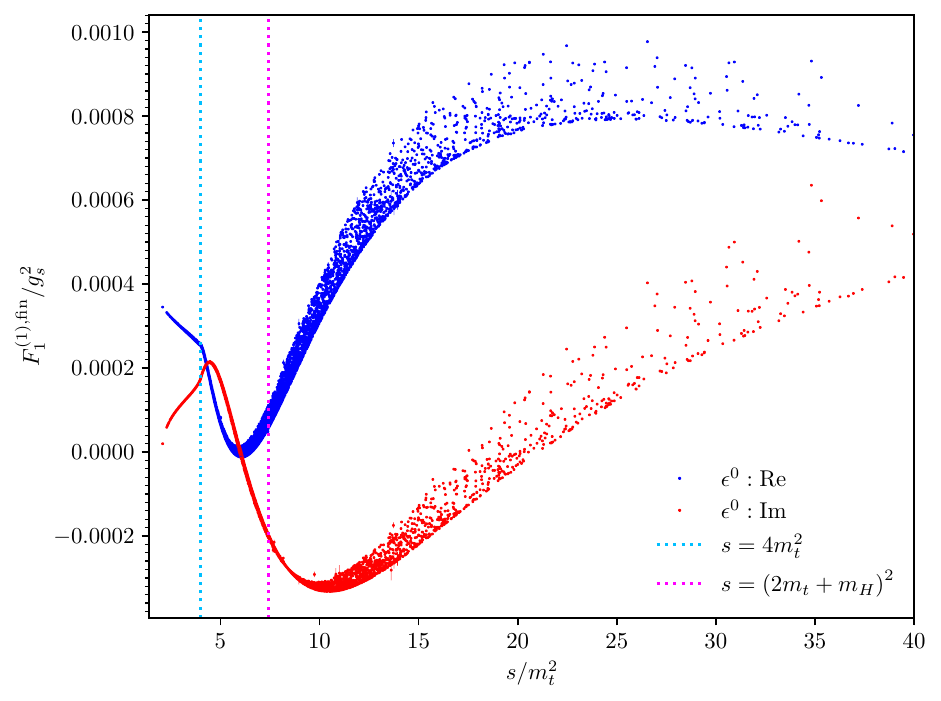}
\end{subfigure}
\begin{subfigure}{0.49\textwidth}
\centering\includegraphics[width=1.0\textwidth]{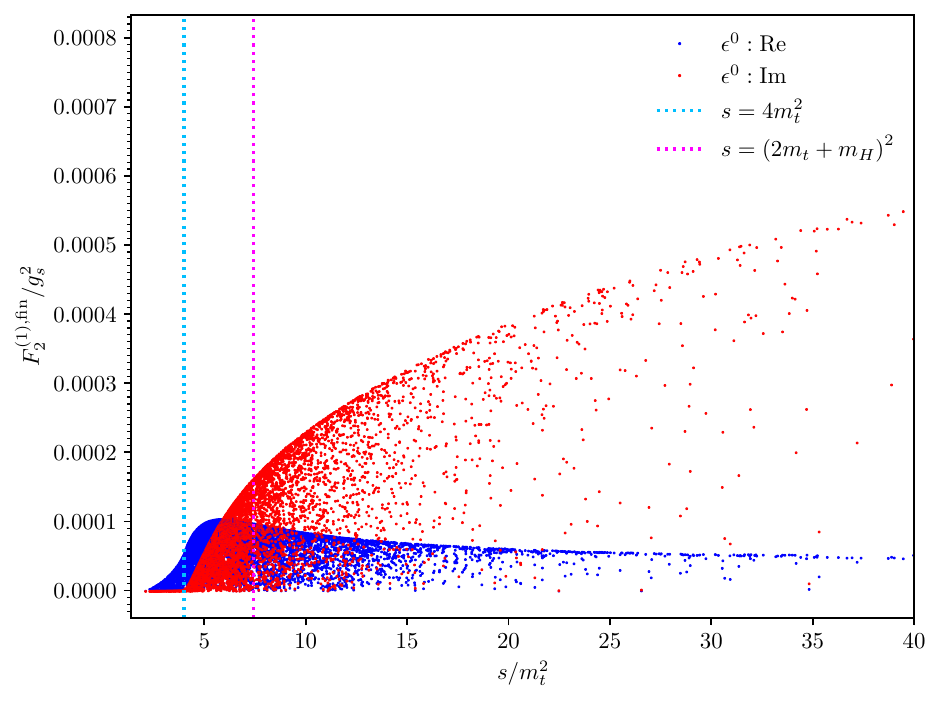}
\end{subfigure}
\caption{The UV-renormalised form factors $F_1^{(1),\mathrm{fin}}$ (left panel) and $F_2^{(1),\mathrm{fin}}$ (right panel) divided by $g_s^2$. 
Note that the spread of points, which is due to the  $t$-dependence, is milder in $F_1^{(1)}$ than in $F_2^{(1)}$.
The uncertainty of each phase-space point due to the limited precision of the numerical integration is indicated with an error bar.
}
\label{fig:ffs}
\end{figure}

In Fig.~\ref{fig:ffs} we display the finite, UV-renormalised, form factors as a function of the Mandelstam invariant $s$.
Examining the $F_1^{(1),\mathrm{fin}}$ form factor we observe that it has both a real and imaginary part for all physically accessible values of $s$, even close to the $HH$ production threshold, this is because it receives a large contribution from diagrams with a two-particle cut through a pair of Higgs bosons (i.e. with a $HH$ threshold), see e.g. Figs.~\ref{sfig:g3g4gt}--\ref{sfig:g32gt2}.
The $t$-dependence, visible in the spread of points at a given $s$ value, is much milder for $F_1^{(1),\mathrm{fin}}$ than $F_2^{(1),\mathrm{fin}}$.
Considering the $F_2^{(1),\mathrm{fin}}$ form factor, we note that it is also complex-valued in the entire physically accessible region of phase-space.
However, only a small imaginary part exists between the $HH$ and $t \overline{t}$ thresholds.
As discussed, the $F_2^{(1),\mathrm{fin}}$ form factor receives contributions only from 1PI diagrams, the only class of diagrams contributing with a $HH$ threshold in the $s$-channel are those of Fig.~\ref{sfig:g32gt2}. 
We find that numerically the contribution of these diagrams to $F_2^{(1),\mathrm{fin}}$ at low invariant mass is much smaller than that of other coupling structures.
In Appendix~\ref{app:bareplots} we present plots of the finite term of the individual bare form factors $F^{(1)}_{i,j}$.

In order to verify our results, we carried out a number of checks. Firstly, we checked that our two independently generated amplitudes (before reduction to masters) were symbolically identical up to sector relations and symmetries. Secondly, we confirmed that the amplitude is symmetric under the exchange of $t$ and $u$ by comparing the numerical results of multiple pairs of phase-space points wherein the first point's $t$-value is substituted by $u=2m_H^{2}-s-t$ in the second and observing that these are identical within the stated numerical error. Thirdly, for the two-loop contribution, we observed that before UV renormalisation the only poles appearing were $1/\epsilon$ (spurious poles up to order $1/\epsilon^{4}$ cancel). After UV renormalisation, all poles cancel which simultaneously corroborates our expectation that there are neither soft nor collinear IR singularities. We also checked that poles of the bare form factors $F_{i,j}^{(1)}$ are purely real below the $t\overline{t}$ threshold for a selection of phase-space points in this kinematic region.

\begin{table}[h!]
\centering
\begin{tabular}{l|l|l|l}
$\sqrt{s}$   & 13 TeV & 13.6 TeV & 14 TeV \\ \hline
$\mathrm{LO}$ {[}fb{]}  & $16.45$   & $18.26$          & $19.52$        \\
$\mathrm{NLO}^\mathrm{EW}$ {[}fb{]} & $16.69$   & $18.52$         & $19.79$       \\ \hline
$\mathrm{NLO}^\mathrm{EW}$/$\mathrm{LO}$       & $1.01$   & $1.01$          & $1.01$       
\end{tabular}
\caption{Inclusive cross section for Higgs boson pair production for different centre-of-mass energies at $\mathrm{LO}$ and $\mathrm{NLO}^\mathrm{EW}$ including only the Yukawa and self-coupling type corrections.
The QCD renormalisation and factorisation scales are set to $\mu_r = \mu_f = m_{HH}/2$.}
\label{tab:xs}
\end{table}

\begin{figure}[h!]
\centering
\begin{subfigure}{0.49\textwidth}
\centering\includegraphics[width=1.0\textwidth]{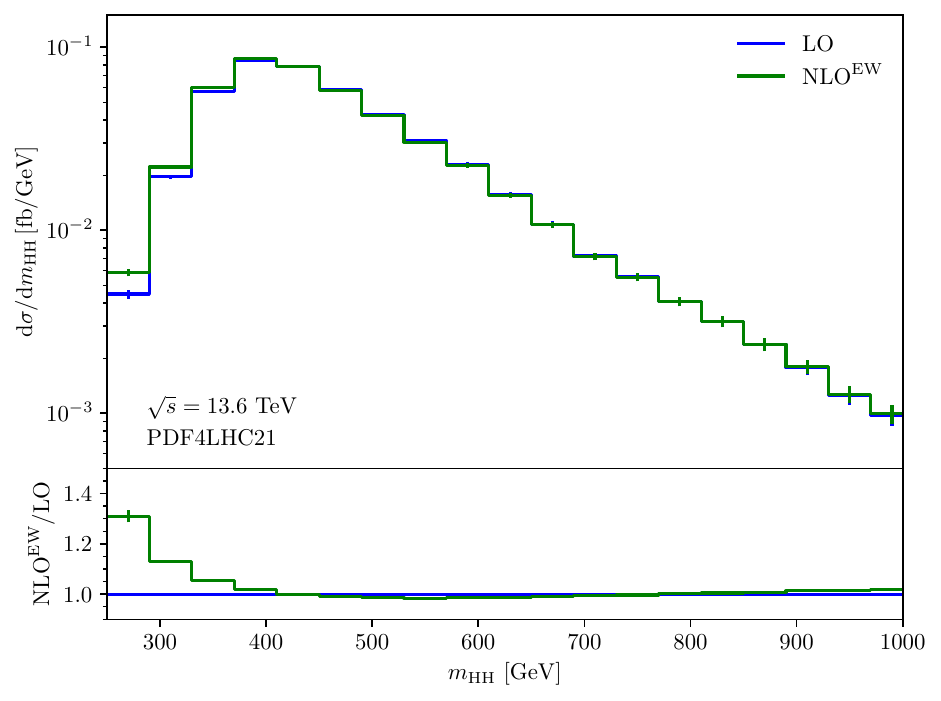}
\end{subfigure}
\begin{subfigure}{0.49\textwidth}
\centering\includegraphics[width=1.0\textwidth]{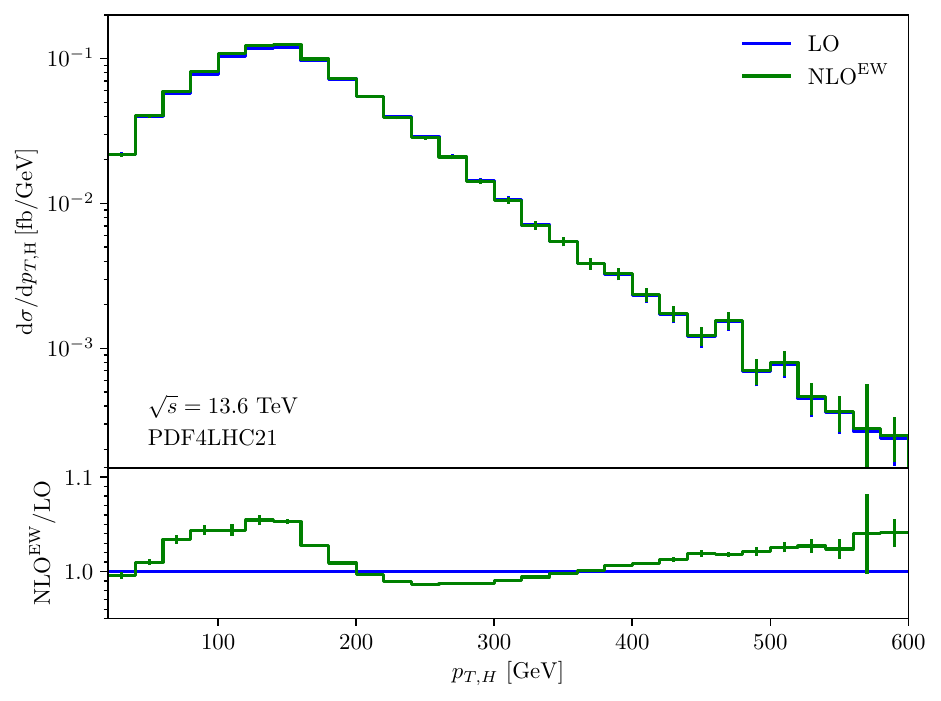}
\end{subfigure}
\caption{Invariant mass and transverse momentum distributions for Higgs boson pair production at LO and $\mathrm{NLO}^\mathrm{EW}$ including only the Yukawa and self-coupling type corrections.
The QCD renormalisation and factorisation scales are set to $\mu_r = \mu_f = m_{HH}/2$.}
\label{fig:dists}
\end{figure}

For the presentation of our final results, we use the \texttt{PDF4LHC21\_40}~\cite{PDF4LHCWorkingGroup:2022cjn} distribution functions interfaced via \textsc{LHAPDF}~\cite{Buckley:2014ana} and set the factorisation and renormalisation scale to $\mu_r = \mu_f = m_{HH}/2$. The masses of the Higgs boson and top quark are set to $m_H = 125~\mathrm{GeV}$, $m_t = \sqrt{23/12}\,m_H =173.055~\mathrm{GeV}$, respectively, and we set $G_F = 1.1663787 \cdot 10^{-5} \ \mathrm{GeV}^{-2}$, corresponding to $v=246.22~\mathrm{GeV}$.

Results for the total and differential cross section at the LHC with a centre-of-mass energy of $\sqrt s = 13\,\mathrm{TeV}, 13.6\,\mathrm{TeV}$ and $14\,\mathrm{TeV}$ are given in Table~\ref{tab:xs} and shown differentially in $m_{HH}$ and $p_{T,H}$ in Fig.~\ref{fig:dists}, respectively.
These results are obtained by reweighting $\sim 7000$ unweighted LO events with the NLO$^\mathrm{EW}$ contribution.
We observe that the partial NLO$^\mathrm{EW}$ corrections computed here increase the total cross section by $\sim 1\%$.
This is comparable to the size of the QCD scale uncertainty of $\sim 3 \%$ obtained at N$^3$LO in the heavy top-quark limit~\cite{Chen:2019lzz,Chen:2019fhs}.

For the invariant mass distribution, shown in Fig.~\ref{fig:dists}, the corrections introduce very large shape distortions, $\sim 30\%$ with the binning we select, close to the Higgs pair production threshold, compatible with the observations of Ref.~\cite{Muhlleitner:2022ijf}.
In Ref.~\cite{Bi:2023bnq}, it was found that the full EW corrections lead to an enhancement of the $m_{HH}$ spectrum close to the Higgs boson pair production threshold of up to 15\%.
Reproducing the binning used in Ref.~\cite{Bi:2023bnq} we find an enhancement of $\sim 25\%$, suggesting that the gauge boson contributions included in their full results partly cancel the enhancement we see at low $m_{HH}$ values.
This appears plausible when looking at individual contributions to the EW corrections for single Higgs boson production~\cite{Aglietti:2004nj,Degrassi:2004mx,Actis:2008ug}.
The shape distortions in the $p_{T,H}$ distribution of our results are less localised, with a significant $5\%$ enhancement just below the top-quark pair production threshold and at high-$p_{T,H}$, along with suppression at the level of $2\%$ just above the top-quark pair production threshold.
In our results, we see a general enhancement at high $m_{HH}$ and $p_{T,H}$ not present in the full EW corrections, this suggests that the gauge boson contribution dominates at high-energy and is negative.

We have also evaluated our results using the \texttt{NNPDF31\_nlo\_as\_0118} PDF set as used in Ref.~\cite{Bi:2023bnq}. Using this PDF set, we obtain a total NLO$^\mathrm{EW}$ cross section of $20.19$ fb including only the Yukawa and self-coupling type corrections, which is a $1\%$ enhancement compared to the LO.
In comparison, the full NLO$^\mathrm{EW}$ total cross section presented in Ref.~\cite{Bi:2023bnq} is $19.12(6)$ fb, which is a $4.2\%$ decrease relative to the LO. 
This discrepancy suggests that the gauge boson contribution, appearing in the complete EW calculation, dominates the corrections and has the opposite sign to the corrections computed here.

\section{Conclusions}
\label{sec:conclusions}
We have presented the calculation of the electroweak corrections to Higgs boson pair production in gluon fusion in the gaugeless limit.
In total, these partial NLO electroweak corrections increase the cross section by about $1\,\%$.
The corrections impact the Higgs boson pair production invariant mass and transverse momentum distributions, giving an enhancement of up to $30\,\%$ at low $m_{HH}$ values due to the Yukawa-type corrections, which is larger than in the case of the full corrections presented in Ref.~\cite{Bi:2023bnq}, where the enhancement is found to be $15\,\%$.
This suggests that the gauge boson contributions are negative for $m_{HH}$ values below the $2m_t$ threshold.
We also observe almost no correction for higher values of $m_{HH}$, in contrast to $-10\,\%$ found in Ref.~\cite{Bi:2023bnq}, suggesting again that this region is dominated by negative contributions from diagrams containing $W$ and $Z$ bosons. 

In our calculation, we retain the full symbolic dependence on the top-quark and Higgs boson masses in the reduction to master integrals of the two-loop amplitude. 
All integrals are calculated using sector decomposition and cross-checked by evaluating them using the series expansion of differential equations. 
We provide results for the bare amplitude divided into individual form factors separated according to the Yukawa, Higgs trilinear and quartic couplings.
We present results for the UV-renormalised form factors, the di-Higgs invariant mass and the Higgs boson transverse momentum distribution. 
The renormalisation of partial electroweak corrections in the Yukawa model is discussed in detail, this provides relevant input for the interpretation of results presented elsewhere in the literature for non-Standard Model values of the Higgs boson self-couplings.

The results presented here, and the techniques used to obtain them, provide an important cross-check and benchmark for further analysing and interpreting the complete electroweak corrections.
For example, the fully symbolic reduction obtained here allows for the study of mass scheme uncertainties.
Our results also facilitate investigating the effects of anomalous couplings, for example, anomalous trilinear and quartic Higgs boson couplings.
These couplings can be varied consistently within an Effective Field Theory framework, for example the
non-linear Effective Field Theory (HEFT), where the fact that the Higgs boson is an EW singlet decorrelates the trilinear and quartic Higgs couplings at leading order in the EFT expansion.
Although not the main focus of this work, our complete set of differential equations, which can be evaluated using series expansion methods, may also provide useful semi-analytic insights into the structure of the electroweak corrections.

\acknowledgments

We thank Vitaly Magerya for support using the \textsc{alibrary} and \textsc{ratracer} packages, Joshua Davies, Go Mishima, Kay Schönwald, Matthias Steinhauser, Hantian Zhang for cross-checking our bare $g_t^4$ form factors with the high-energy expansion obtained in Ref.~\cite{Davies:2022ram}, 
Martijn Hidding for support with (and providing a modified version of) \textsc{DiffExp},
and Benjamin Pecjak, Tommy Smith, Max Löschner for clarifying discussions regarding input-parameter schemes and renormalisation.

This research was supported in part by the Deutsche Forschungsgemeinschaft (DFG, German Research Foundation) under grant 396021762 - TRR 257, and by the UK Science and Technology Facilities Council under contract ST/T001011/1. 
SJ is supported by a Royal Society University Research Fellowship (Grant URF/R1/201268).
GH, SJ, TS and AV are grateful to the Galileo Galilei Institute for hospitality and support during the scientific program on ``Theory Challenges in the Precision Era of the Large Hadron Collider'', where part of this work was done.
GH would like to thank the Institute for Particle Physics Phenomenology at Durham University for hospitality while part of this work was carried out.
\clearpage

\appendix

\section{Details of Renormalisation}
\label{app:renormalisation}
From the Lagrangian of Eq.~\eqref{eq:LLO} one arrives at a fully renormalised theory by first including the vev shift $ v_0 \rightarrow v_0 + \Delta v $ to obtain
\begin{align}
	\begin{split}\mathcal{L}'=&\frac{1}{2}(\partial_\mu H_0)^\dagger(\partial^\mu H_0) + \frac{\mu^2_0}{2} (v_0 + \Delta v + H_0)^2 + \frac{\lambda_0}{16} (v_0 + \Delta v + H_0)^4\\
    &+ i \bar t_0 \slashed D t_0 - y_{t,0} \frac{v_0 + \Delta v + H_0}{\sqrt{2}}\bar t_0t_0\end{split}\\
	\begin{split}=&\frac{1}{2}(\partial_\mu H_0)^\dagger(\partial^\mu H_0) + H_0 \left(\mu_0^2 v_0 + \frac{\lambda_0 v_0^3}{4} + \Delta v (\mu^2_0 + \frac{3}{4}\lambda_0 v_0^2)\right)\\
    &+ H_0^2 \left(\frac{\mu_0^2}{2} + \frac{3 v_0^2 \lambda_0}{8} + \frac{3}{4}\lambda_0 v_0 \Delta v\right) + H_0^3 \left(\frac{\lambda_0 v_0}{4} + \Delta v \frac{\lambda_0}{4}\right) + H_0^4 \frac{\lambda_0}{16}\\
    &+ i\bar t_0\slashed D t_0 - m_{t,0}\bar t_0t_0 - \frac{m_{t,0}}{v_0}\Delta v \bar t_0t_0 - \frac{m_{t,0}}{v_0}H_0\bar t_0t_0~.\label{eq:Lvevshift}\end{split}
\end{align}
This step is required to keep the value of $ v_0 $ at the minimum of the Higgs potential, which is shifted at NLO compared to LO. On a diagrammatic level, the shift of the minimum of the Higgs potential is caused by diagrams containing tadpole sub-diagrams. 

The definition of the vev upon renormalisation is therefore related to the treatment of tadpole contributions. Tadpole counterterms can be generated in two different ways in the Lagrangian: through parameter renormalisation~\cite{Denner:1991kt,Denner:2016etu,Denner:2018opp}, or via Higgs field redefinitions~\cite{Fleischer:1980ub,Actis:2006ra,Krause:2016oke,Denner:2016etu}, see Ref.~\cite{Denner:2019vbn} for a review. The latter is also called Fleischer-Jegerlehner scheme (FJTS).
A new scheme for tadpole renormalisation, dubbed Gauge-Invariant Vacuum expectation value Scheme (GIVS), has been suggested recently~\cite{Dittmaier:2022maf}, which is a hybrid scheme of the two schemes mentioned above, with the benefits of being gauge independent while avoiding large corrections in $\overline{\mathrm{MS}}$-type schemes. The effects of certain input parameter schemes in SMEFT have been studied in Ref.~\cite{Biekotter:2023xle}.

When using OS renormalisation in unitary gauge, the FJTS is a suitable choice yielding the vev shift prescription and thereby the Lagrangian of Eq.\eqref{eq:Lvevshift}. 
The emerging term linear in the Higgs field  is identified with the tadpole counterterm
\begin{equation}\label{eq:deltaT}
	\delta T= \left(\mu_0^2 v_0 + \frac{\lambda_0 v_0^3}{4} + \Delta v (\mu^2_0 + \frac{3}{4}\lambda_0 v_0^2)\right) = -\Delta v m_H^2~,
\end{equation}
where the first two terms in the brackets cancel upon using Eq.~\eqref{eq:mu_vev_mh_relation} and, in the second equality, the bare quantities have been expressed in terms of their renormalised counterparts, neglecting higher-order terms of $\mathcal{O}(\delta_X^2, \Delta v \delta_X, (\Delta v)^2 )$. 
The renormalisation condition is that the sum of the tadpoles, $T^H$, and the tadpole counterterm, $\delta T$, should vanish at the given order,
\begin{equation}\label{eq:deltaT2}
	0\overset{!}{=}\delta T + T^H\quad\Leftrightarrow\quad\delta T = -T^H = -\big[\raisebox{-2.5ex}{\raisebox{0.5ex}{\scalebox{0.3}{\begin{tikzpicture}
	\begin{pgfonlayer}{nodelayer}
		\node [style=none] (1) at (-2, 0) {};
		\node [style=none] (2) at (2, 0) {};
		\node [style=dot] (3) at (0, 0) {};
	\end{pgfonlayer}
	\begin{pgfonlayer}{edgelayer}
		\draw [style=scalar] (1.center) to (3);
		\draw [style=scalar, bend left=90, looseness=1.75] (3) to (2.center);
		\draw [style=scalar, bend right=90, looseness=1.75] (3) to (2.center);
	\end{pgfonlayer}
\end{tikzpicture}}}}+\raisebox{-2.5ex}{\raisebox{0.5ex}{\scalebox{0.3}{\begin{tikzpicture}
	\begin{pgfonlayer}{nodelayer}
		\node [style=none] (1) at (-2, 0) {};
		\node [style=none] (2) at (2, 0) {};
		\node [style=dot] (3) at (0, 0) {};
	\end{pgfonlayer}
	\begin{pgfonlayer}{edgelayer}
		\draw [style=scalar] (1.center) to (3);
		\draw [style=fermionarrow] (2.center)
			 to [bend left=90, looseness=1.75] (3.center)
			 to [bend left=90, looseness=1.75] cycle;
	\end{pgfonlayer}
\end{tikzpicture}}}}\big]~.
\end{equation}
With this condition, all contributions from tadpole subdiagrams are integrated out and collected in the counterterm $ \delta T $.
Inserting Eq.~\eqref{eq:deltaT} as well as the field, parameter, and vertex renormalisations from Eqs.~\eqref{eq:H0} through \eqref{eq:vev0} into Eq.~\eqref{eq:Lvevshift} yields the fully renormalised Lagrangian
\begin{align}
	\begin{split}\label{eq:Lrenormalised}
	\mathcal{L}=&\frac{1}{2}(1+\delta_H)(\partial_\mu H)^\dagger(\partial^\mu H) + H \delta T - \left(\frac{m_H^2}{2}\left(1+\delta m_H^2+\delta_H\right) - \frac{3\delta T}{2v}\right) H^2\\
	&- \left(\frac{g_3}{3!}\left(1+\delta m_H^2 + \frac{3}{2}\delta_H - \delta_v\right) - \frac{\delta T}{2v^2}\right) H^3 - \frac{g_4}{4!}\left(1+\delta m_H^2 + 2\delta_H - 2\delta_v\right) H^4\\
	&+ i(1+\delta_t)\bar t\slashed D t - m_t\left(1+\delta m_t + \delta_t-\frac{\delta T}{vm_H^2}\right)\bar tt -  g_t\left(1+\delta m_t+\frac{\delta_H}{2}+\delta_t - \delta_v\right)H\bar tt,
	\end{split}
\end{align}
where the couplings $g_3, g_4$ and $g_t$ are the renormalised counterparts of the bare couplings, $g_{3,0}, g_{4,0}$ and $g_{t,0}$. 
They obey Eq.~\eqref{eq:SMcouplings} after substituting the bare quantities with their renormalised values.
Since the explicit tadpole insertions into each diagram now cancel with the corresponding explicit tadpole counterterm insertions, we can neglect both of these explicit contributions.
Tadpole contributions will therefore only appear implicitly due to the terms $\delta T$ appearing in the counterterm insertions, given in Appendix~\ref{app:rulesExpr} (see also Ref.~\cite{Denner:2019vbn} Section 3.1.7, where they use the notation $\delta t$ to denote what we call $\delta T$ in the present work).

We perform an on-shell renormalisation, which fixes $ \delta_H $, $ \delta_t $, $ \delta m_H^2 $ and $ \delta m_t $ via the renormalisation conditions
\begin{align}\label{eq:onshellcond}
	0 = \Bigl[\Sigma(\slashed p)\Bigr]_{\slashed p=m}\;\; , \;\;
	0 = \Biggl[\frac{\mathrm{d}}{\mathrm{d}\slashed p}\Sigma(\slashed p)\Biggr]_{\slashed p = m}~.
\end{align}
The masses $ m $ and self-energies $ \Sigma $ are those of the top quark and the Higgs boson, respectively. For the top self-energy $\Sigma_t$, only the mixed top-Higgs bubble \raisebox{-2ex}{\raisebox{0.5ex}{\scalebox{0.3}{\begin{tikzpicture}
	\begin{pgfonlayer}{nodelayer}
		\node [style=none] (1) at (-2, 0) {};
		\node [style=none] (2) at (2, 0) {};
		\node [style=dot] (3) at (0, 0) {};
        \node [style=none] (4) at (4, 0) {};
	\end{pgfonlayer}
	\begin{pgfonlayer}{edgelayer}
		\draw [style=fermionarrow] (1.center) to (3);
		\draw [style=fermionarrow, bend left=90, looseness=1.4] (3.center) to (2.center);
        \draw [style=scalar, bend left=90, looseness=1.4] (2.center) to (3.center);
        \draw [style=fermionarrow] (2.center) to (4);
	\end{pgfonlayer}
\end{tikzpicture}}}} and the counterterm insertion \raisebox{-0.3ex}{\raisebox{0.5ex}{\scalebox{0.3}{\begin{tikzpicture}
	\begin{pgfonlayer}{nodelayer}
		\node [style=none] (0) at (-2, 0) {};
		\node [style=none] (1) at (2, 0) {};
	\end{pgfonlayer}
	\begin{pgfonlayer}{edgelayer}
		\draw [style=ct fermion] (0.center) to (1.center);
	\end{pgfonlayer}
\end{tikzpicture}
}}} contribute whereas for the Higgs self-energy $\Sigma_H$, there are three diagrams and the counterterm insertion. The resulting renormalisation constants are given in Appendix~\ref{app:rulesExpr}.

The vev counterterm can be fixed using any of the Yukawa, triple, or quartic Higgs self-interaction vertices. 
For consistency with much of the literature on EW corrections, we employ the $G_\mu$ scheme and use the counterterm as given in Ref.~\cite{Biekotter:2023xle} in the limit $M_W \rightarrow 0, M_Z \rightarrow 0$, as detailed in Section~\ref{ssec:dvev}.

Finally, we note that the top-quark wave function renormalisation counterterm $\delta_t$ enters in multiple vertices, but since there are only closed top loops occurring, the final result should not contain any dependence on this quantity. 
Every top vertex counterterm insertion $\propto \delta_t$ is cancelled by the top propagator insertion $\propto \delta_t^{-1}$. This also serves as a crosscheck of the renormalised amplitude and, indeed, we do not observe any dependence on $\delta_t$ in our final expression.

\subsection{Vacuum Expectation Value Counterterm}
\label{ssec:dvev}
\begin{figure}
\centering
\begin{subfigure}{0.3\textwidth}
\centering
\begin{tikzpicture}[scale=0.8]
	\begin{pgfonlayer}{nodelayer}
		\node [style=none] (0) at (-2, -1.25) {};
		\node [style=none] (1) at (2, -1.25) {};
		\node [style=none] (2) at (0, 1.75) {};
		\node [style=dot] (3) at (0, 0.6) {};
            \node [style=dot] (4) at (0.75, -0.65) {};
            \node [style=dot] (5) at (-0.75, -0.65) {};
	\end{pgfonlayer}
	\begin{pgfonlayer}{edgelayer}
		\draw [style=scalar] (0) to (5);
		\draw [style=scalar] (5) to (4);
		\draw [style=scalar] (4) to (1);
            \draw [style=scalar] (3) to (4);
            \draw [style=scalar] (3) to (2);
            \draw [style=scalar] (5) to (3);
	\end{pgfonlayer}
\end{tikzpicture}
\caption{}\label{sfig:a}
\end{subfigure}
\hfill
\begin{subfigure}{0.3\textwidth}
\centering
\begin{tikzpicture}[scale=0.8]
	\begin{pgfonlayer}{nodelayer}
		\node [style=none] (0) at (-2, -1.25) {};
		\node [style=none] (1) at (2, -1.25) {};
		\node [style=none] (2) at (0, 1.75) {};
		\node [style=dot] (3) at (0, 0.6) {};
            \node [style=dot] (4) at (0.75, -0.65) {};
            \node [style=dot] (5) at (-0.75, -0.65) {};
	\end{pgfonlayer}
	\begin{pgfonlayer}{edgelayer}
		\draw [style=scalar] (0) to (5);
		\draw [style=fermionarrow] (4) to (5);
		\draw [style=scalar] (4) to (1);
            \draw [style=fermionarrow] (3) to (4);
            \draw [style=scalar] (3) to (2);
            \draw [style=fermionarrow] (5) to (3);
	\end{pgfonlayer}
\end{tikzpicture}
\caption{}\label{sfig:b}
\end{subfigure}
\hfill
\begin{subfigure}{0.3\textwidth}
\centering
\begin{tikzpicture}[scale=0.8]
	\begin{pgfonlayer}{nodelayer}
		\node [style=none] (0) at (-2, -1.25) {};
		\node [style=none] (1) at (2, -1.25) {};
		\node [style=none] (2) at (0, 1.75) {};
            \node [style=dot] (10) at (0, -0.3) {};
		\node [style=dot] (3) at (0, 0.6) {};
	\end{pgfonlayer}
	\begin{pgfonlayer}{edgelayer}
		\draw [style=scalar] (0) to (10);
		\draw [style=scalar] (1) to (10);
            \draw [style=scalar] (3) to (2);
            \draw [style=scalar] (10) arc (-90:270:0.45);;
	\end{pgfonlayer}
\end{tikzpicture}
\caption{}\label{sfig:c}
\end{subfigure}\\
\begin{subfigure}{0.3\textwidth}
\centering
\begin{tikzpicture}[scale=0.8]
	\begin{pgfonlayer}{nodelayer}
		\node [style=none] (0) at (-2, -1.25) {};
		\node [style=none] (1) at (2, -1.25) {};
		\node [style=none] (2) at (0, 1.75) {};
		\node [style=dot] (3) at (0, 0.6) {};
            \node [style=dot] (4) at (0.75, -0.65) {};
            \node [style=dot] (5) at (-0.75, -0.65) {};
	\end{pgfonlayer}
	\begin{pgfonlayer}{edgelayer}
		\draw [style=fermionarrow] (0) to (5);
		\draw [style=scalar] (5) to (4);
		\draw [style=fermionarrow] (4) to (1);
            \draw [style=fermionarrow] (3) to (4);
            \draw [style=scalar] (2) to (3);
            \draw [style=fermionarrow] (5) to (3);
	\end{pgfonlayer}
\end{tikzpicture}
\caption{}\label{sfig:d}
\end{subfigure}\hfill
\begin{subfigure}{0.3\textwidth}
\centering
\begin{tikzpicture}[scale=0.8]
	\begin{pgfonlayer}{nodelayer}
		\node [style=none] (0) at (-2, -1.25) {};
		\node [style=none] (1) at (2, -1.25) {};
		\node [style=none] (2) at (0, 1.75) {};
		\node [style=dot] (3) at (0, 0.6) {};
            \node [style=dot] (4) at (0.75, -0.65) {};
            \node [style=dot] (5) at (-0.75, -0.65) {};
	\end{pgfonlayer}
	\begin{pgfonlayer}{edgelayer}
		\draw [style=fermionarrow] (0) to (5);
		\draw [style=fermionarrow] (5) to (4);
		\draw [style=fermionarrow] (4) to (1);
            \draw [style=scalar] (3) to (4);
            \draw [style=scalar] (3) to (2);
            \draw [style=scalar] (5) to (3);
	\end{pgfonlayer}
\end{tikzpicture}
\caption{}\label{sfig:e}
\label{fig:vevsubfig}
\end{subfigure}\hfill
\begin{subfigure}{0.3\textwidth}
\centering
\begin{tikzpicture}[scale=0.8]
	\begin{pgfonlayer}{nodelayer}
		\node [style=none] (0) at (-2, -1.25) {};
		\node [style=none] (1) at (2, -1.25) {};
		\node [style=none] (2) at (2, 1.75) {};
		\node [style=none] (3) at (-2, 1.75) {};
            \node [style=dot] (4) at (1, -0.65) {};
            \node [style=dot] (5) at (-1, -0.65) {};
            \node [style=dot] (6) at (-1, 1.05) {};
            \node [style=dot] (7) at (1, 1.05) {};
	\end{pgfonlayer}
	\begin{pgfonlayer}{edgelayer}
		\draw [style=scalar] (0) to (5);
		\draw [style=scalar] (5) to (4);
		\draw [style=scalar] (4) to (1);
		\draw [style=scalar] (3) to (6);
		\draw [style=scalar] (6) to (7);
		\draw [style=scalar] (7) to (2);
		\draw [style=scalar] (6) to (5);
		\draw [style=scalar] (7) to (4);
	\end{pgfonlayer}
\end{tikzpicture}
\caption{}\label{sfig:f}
\end{subfigure}\\
\begin{subfigure}{0.3\textwidth}
\centering
\begin{tikzpicture}[scale=0.8]
	\begin{pgfonlayer}{nodelayer}
		\node [style=none] (0) at (-2, -1.25) {};
		\node [style=none] (1) at (2, -1.25) {};
		\node [style=none] (2) at (2, 1.75) {};
		\node [style=none] (3) at (-2, 1.75) {};
            \node [style=dot] (4) at (1, -0.65) {};
            \node [style=dot] (5) at (-1, -0.65) {};
            \node [style=dot] (6) at (-1, 1.05) {};
            \node [style=dot] (7) at (1, 1.05) {};
	\end{pgfonlayer}
	\begin{pgfonlayer}{edgelayer}
		\draw [style=scalar] (0) to (5);
		\draw [style=fermionarrow] (5) to (4);
		\draw [style=scalar] (4) to (1);
		\draw [style=scalar] (3) to (6);
		\draw [style=fermionarrow] (7) to (6);
		\draw [style=scalar] (7) to (2);
		\draw [style=fermionarrow] (6) to (5);
		\draw [style=fermionarrow] (4) to (7);
	\end{pgfonlayer}
\end{tikzpicture}
\caption{}\label{sfig:g}
\end{subfigure}\hfill
\begin{subfigure}{0.3\textwidth}
\centering
\begin{tikzpicture}[scale=0.8]
	\begin{pgfonlayer}{nodelayer}
		\node [style=none] (0) at (-2, -1.25) {};
		\node [style=none] (1) at (2, -1.25) {};
		\node [style=none] (2) at (2, 1.75) {};
		\node [style=none] (3) at (-2, 1.75) {};
            \node [style=dot] (4) at (1, 0.25) {};
            \node [style=dot] (5) at (-1, 0.25) {};
            \node [style=dot] (6) at (0, 0.25) {};
	\end{pgfonlayer}
	\begin{pgfonlayer}{edgelayer}
		\draw [style=scalar] (0) to (5);
		\draw [style=scalar] (3) to (5);
		\draw [style=scalar] (6) to (4);
		\draw [style=scalar] (4) to (1);
		\draw [style=scalar] (4) to (2);
		\draw [style=scalar] (5) arc (180:540:0.5);
	\end{pgfonlayer}
\end{tikzpicture}
\caption{}\label{sfig:h}
\end{subfigure}\hfill
\begin{subfigure}{0.3\textwidth}
\centering
\begin{tikzpicture}[scale=0.8]
	\begin{pgfonlayer}{nodelayer}
		\node [style=none] (0) at (-2, -1.25) {};
		\node [style=none] (1) at (2, -1.25) {};
		\node [style=none] (2) at (2, 1.75) {};
		\node [style=none] (3) at (-2, 1.75) {};
            \node [style=dot] (4) at (1, 0.25) {};
            \node [style=dot] (5) at (-1, 0.25) {};
	\end{pgfonlayer}
	\begin{pgfonlayer}{edgelayer}
		\draw [style=scalar] (0) to (5);
		\draw [style=scalar] (3) to (5);
		\draw [style=scalar] (4) to (1);
		\draw [style=scalar] (4) to (2);
		\draw [style=scalar] (5) arc (180:540:1);
	\end{pgfonlayer}
\end{tikzpicture}
\caption{}\label{sfig:i}
\end{subfigure}
\caption{Example diagrams contributing to the fixing of $\delta_v$ from the Higgs cubic vertex (\subref{sfig:a}, \subref{sfig:b}, \subref{sfig:c}), the Yukawa vertex (\subref{sfig:d}, \subref{sfig:e}) and the Higgs quartic vertex (\subref{sfig:f}, \subref{sfig:g}, \subref{sfig:h}, \subref{sfig:i}).}
\label{fig:1loopdiags}
\end{figure}
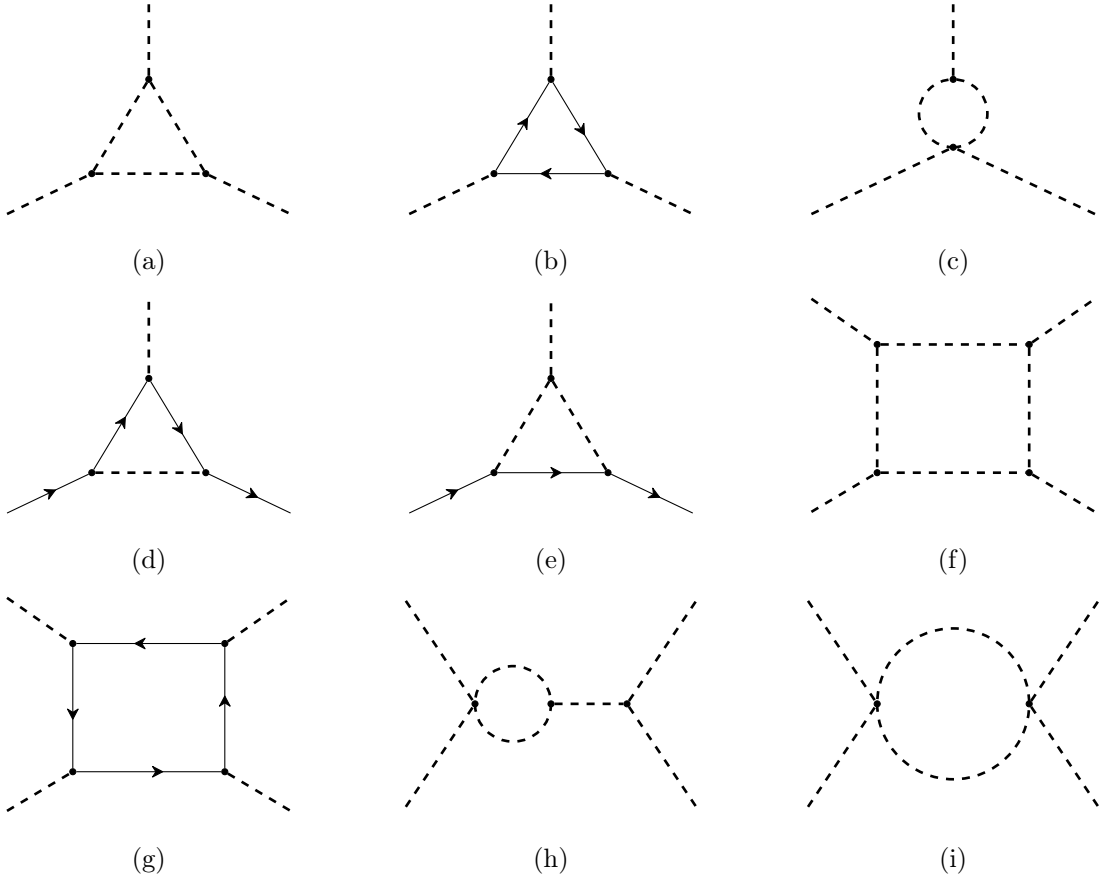
The vev counterterm can be fixed by demanding the finiteness of the higher-order electroweak corrections to an electroweak vertex of the theory. 
For determining the poles of the counterterm, it does not matter which vertex is picked, and we are free to use either the Yukawa, Higgs cubic or Higgs quartic vertex. 
By explicit calculation, we find that all three vertices give the same UV divergent part for the vev counterterm,
\begin{equation}\label{eq:dvUV}
	\delta_v|_\mathrm{UV}=-\frac{3m_H^4+2m_H^2m_t^2N_c-8m_t^4N_c}{32\pi^2m_H^2v^2\epsilon},
\end{equation}
upon demanding that NLO electroweak virtual contributions do not correct the tree-level expression for the vertex. 
For example, for the Yukawa coupling, we may require that
\begin{equation}
	- i g_t \overset{!}{=} \Gamma_{H\bar tt},
\end{equation}
at NLO with the diagrams in Figs.~\ref{sfig:d} and \ref{sfig:e} contributing to $\Gamma_{H\bar tt}$.
This fixes the vev counterterm and we find that the divergent part in our theory is given by
\begin{equation}\label{eq:dvUVyukawa}
\delta_v^{g_t}\!\left(g_t,g_3,g_4\right)\!|_\mathrm{UV}=-\frac{g_3 g_t m_H^2+2g_t^2m_t\left(m_H^2-4m_t^2\right)N_c}{32\pi^2m_H^2m_t\epsilon}
\end{equation}
where $g_4$ is included as an argument because $\delta_v^{g_t}$ can, in principle, have a $g_4$-dependence at higher-orders (but the UV part at NLO explicitly does not). If instead we fix the vev counterterm from the Higgs cubic self-coupling by requiring that
\begin{equation}
	- i g_3 \overset{!}{=} \Gamma_{HHH}
\end{equation}
holds to NLO in our theory -- contributions include the diagrams in Figs.~\ref{sfig:a}, \ref{sfig:b} and \ref{sfig:c}  -- then we obtain,
\begin{equation}
\begin{split}\label{eq:dvUVcubic}
	\delta_v^{g_3}\!\left(g_t,g_3,g_4\right)\!|_\mathrm{UV}=&-\frac{1}{32\pi^2g_3m_H^4\epsilon}\Bigl[g_3g_4m_H^4+8g_4g_tm_H^2m_t^3N_c-8g_3^2g_tm_t^3N_c\\&+2g_3g_t^2m_H^2\left(m_H^2+12m_t^2\right)N_c-48g_t^3m_H^4m_tN_c\Bigr].
\end{split}
\end{equation}
Similarly, from the requirement
\begin{equation}
    - i g_4 \overset{!}{=} \Gamma_{HHHH}~,
\end{equation}
we obtain
\begin{equation}\label{eq:dvUVquartic}
\delta_v^{g_4}\!\left(g_t,g_3,g_4\right)\!|_\mathrm{UV}=-\frac{2g_tg_4N_c(g_t(m_H^4+6m_H^2m_t^2) - 2g_3m_t^3)+g_4^2m_H^4-24g_t^4m_H^4N_c}{32\pi^2g_4m_H^4\epsilon}
\end{equation}
whose derivation includes contributions from the diagrams in Figs.~\ref{sfig:f}, \ref{sfig:g}, \ref{sfig:h} and \ref{sfig:i}.
Upon insertion of the SM coupling values of Eq.~\eqref{eq:SMcouplings}, all of our calculations of the divergent part of $\delta_v$ coincide. 
That is to say,
\begin{equation}\label{eq:dvUVequality}
	\delta_v^{g_t}\!\left(\frac{m_t}{v},\frac{3m_H^2}{v},\frac{3m_H^2}{v^2}\right)\!\Bigg|_\mathrm{UV}\hspace{-3mm}=\delta_v^{g_3}\!\left(\frac{m_t}{v},\frac{3m_H^2}{v},\frac{3m_H^2}{v^2}\right)\!\Bigg|_\mathrm{UV}\hspace{-3mm}=\delta_v^{g_4}\!\left(\frac{m_t}{v},\frac{3m_H^2}{v},\frac{3m_H^2}{v^2}\right)\!\Bigg|_\mathrm{UV}\hspace{-3mm}\overset{!}{=}\delta_v|_\mathrm{UV}
\end{equation}
as they must since the pole cancellation has to occur independently of the scheme choice.
The finite terms, on the other hand, differ.
To obtain a result comparable with other authors' works, we choose the $G_\mu$ scheme. The pole structure also agrees in this case; to fix the finite part we use the result of Ref.~\cite{Biekotter:2023xle} which is obtained from the full SM contributions in $G_\mu$ scheme. After application of the same limits as in the Lagrangian, namely $M_W,M_Z\to0$, we arrive at the counterterm in Eq.~\eqref{eq:dvev}.

\subsection{Feynman Rules \& Counterterm Expressions}
\label{app:rulesExpr}
The Lagrangian of Eq.~\eqref{eq:Lrenormalised} yields the Feynman rules for renormalised quantities and the counterterm insertions.
\begin{table}[h!]
\centering
\begin{tabular}{llp{1cm}ll}
    \vspace{0.2cm} 
    \raisebox{-0.5ex}{\raisebox{0.5ex}{\scalebox{0.3}{\begin{tikzpicture}
	\begin{pgfonlayer}{nodelayer}
		\node [style=none] (0) at (-2, 0) {};
		\node [style=none] (1) at (2, 0) {};
	\end{pgfonlayer}
	\begin{pgfonlayer}{edgelayer}
		\draw [style=scalar] (0.center) to (1.center);
	\end{pgfonlayer}
\end{tikzpicture}}}} & $\frac{i}{p^2-m_H^2}$ & & \raisebox{-0.5ex}{\raisebox{0.5ex}{\scalebox{0.3}{\begin{tikzpicture}
	\begin{pgfonlayer}{nodelayer}
		\node [style=none] (0) at (-2, 0) {};
		\node [style=none] (1) at (2, 0) {};
	\end{pgfonlayer}
	\begin{pgfonlayer}{edgelayer}
		\draw [style=ct scalar] (0.center) to (1.center);
	\end{pgfonlayer}
\end{tikzpicture}}}} & $-i\left[(m_H^2 - p^2)\delta_H + m_H^2\delta m_H^2 - \frac{g_3}{m_H^2}\delta T\right]$ \\
    \raisebox{-0.5ex}{\raisebox{0.5ex}{\scalebox{0.3}{\begin{tikzpicture}
	\begin{pgfonlayer}{nodelayer}
		\node [style=none] (0) at (-2, 0) {};
		\node [style=none] (1) at (2, 0) {};
	\end{pgfonlayer}
	\begin{pgfonlayer}{edgelayer}
		\draw [style=fermionarrow] (0.center) to (1.center);
	\end{pgfonlayer}
\end{tikzpicture}}}} & $\frac{i(\slashed p + m_t)}{p^2 - m_t^2}$ & & \raisebox{-0.5ex}{\raisebox{0.5ex}{\scalebox{0.3}{\input{figures/ctt.tikz}}}} & $-i\left[(m_t-\slashed p)\delta_t + m_t\delta m_t - \frac{g_t}{m_H^2}\delta T\right]$ \\
    \raisebox{-2.5ex}{\raisebox{0.5ex}{\scalebox{0.3}{\begin{tikzpicture}
	\begin{pgfonlayer}{nodelayer}
		\node [style=none] (0) at (-2, -1.25) {};
		\node [style=none] (1) at (2, -1.25) {};
		\node [style=none] (2) at (0, 1.75) {};
		\node [style=dot] (3) at (0, 0) {};
	\end{pgfonlayer}
	\begin{pgfonlayer}{edgelayer}
		\draw [style=gluoncoil] (2.center) to (3);
		\draw [style=fermionarrow] (0.center) to (3);
		\draw [style=fermionarrow] (3) to (1.center);
	\end{pgfonlayer}
\end{tikzpicture}}}} & $i g_s t^a \gamma_\mu$ & & \raisebox{-2.5ex}{\raisebox{0.5ex}{\scalebox{0.3}{\begin{tikzpicture}
	\begin{pgfonlayer}{nodelayer}
		\node [style=none] (0) at (-2, -1.25) {};
		\node [style=none] (1) at (2, -1.25) {};
		\node [style=none] (2) at (0, 1.75) {};
		\node [style=star] (3) at (0, 0) {};
	\end{pgfonlayer}
	\begin{pgfonlayer}{edgelayer}
		\draw [style=gluoncoil] (2.center) to (3);
		\draw [style=fermionarrow] (0.center) to (3);
		\draw [style=fermionarrow] (3) to (1.center);
	\end{pgfonlayer}
\end{tikzpicture}}}} & $ i g_s \delta_t t^a \gamma_\mu$ \\
    \raisebox{-2.5ex}{\raisebox{0.5ex}{\scalebox{0.3}{\begin{tikzpicture}
	\begin{pgfonlayer}{nodelayer}
		\node [style=none] (0) at (-2, -1.25) {};
		\node [style=none] (1) at (2, -1.25) {};
		\node [style=none] (2) at (0, 1.75) {};
		\node [style=dot] (3) at (0, 0) {};
	\end{pgfonlayer}
	\begin{pgfonlayer}{edgelayer}
		\draw [style=scalar] (2.center) to (3);
		\draw [style=fermionarrow] (0.center) to (3);
		\draw [style=fermionarrow] (3) to (1.center);
	\end{pgfonlayer}
\end{tikzpicture}}}} & $- ig_t$ & & \raisebox{-2.5ex}{\raisebox{0.5ex}{\scalebox{0.3}{\begin{tikzpicture}
	\begin{pgfonlayer}{nodelayer}
		\node [style=none] (0) at (-2, -1.25) {};
		\node [style=none] (1) at (2, -1.25) {};
		\node [style=none] (2) at (0, 1.75) {};
		\node [style=star] (3) at (0, 0) {};
	\end{pgfonlayer}
	\begin{pgfonlayer}{edgelayer}
		\draw [style=scalar] (2.center) to (3);
		\draw [style=fermionarrow] (0.center) to (3);
		\draw [style=fermionarrow] (3) to (1.center);
	\end{pgfonlayer}
\end{tikzpicture}}}} & $- ig_t\left(\delta m_t + \frac{\delta_H}{2} + \delta_t - \delta_v\right)$ \\
    \raisebox{-2.5ex}{\raisebox{0.5ex}{\scalebox{0.3}{\begin{tikzpicture}
	\begin{pgfonlayer}{nodelayer}
		\node [style=none] (0) at (-2, -1.25) {};
		\node [style=none] (1) at (2, -1.25) {};
		\node [style=none] (2) at (0, 1.75) {};
		\node [style=dot] (3) at (0, 0) {};
	\end{pgfonlayer}
	\begin{pgfonlayer}{edgelayer}
		\draw [style=scalar] (0.center) to (3);
		\draw [style=scalar] (3) to (1.center);
		\draw [style=scalar] (2.center) to (3);
	\end{pgfonlayer}
\end{tikzpicture}}}} & $-ig_3$ & & \raisebox{-2.5ex}{\raisebox{0.5ex}{\scalebox{0.3}{\begin{tikzpicture}
	\begin{pgfonlayer}{nodelayer}
		\node [style=none] (0) at (-2, -1.25) {};
		\node [style=none] (1) at (2, -1.25) {};
		\node [style=none] (2) at (0, 1.75) {};
		\node [style=star] (3) at (0, 0) {};
	\end{pgfonlayer}
	\begin{pgfonlayer}{edgelayer}
		\draw [style=scalar] (0.center) to (3);
		\draw [style=scalar] (3) to (1.center);
		\draw [style=scalar] (2.center) to (3);
	\end{pgfonlayer}
\end{tikzpicture}}}} & $- ig_3\left(\delta m_H^2 + \frac{3}{2}\delta_H - \delta_v\right) + i \frac{g_4}{m_H^2}\delta T$ \\
    \raisebox{-4ex}{\raisebox{0.5ex}{\scalebox{0.3}{\begin{tikzpicture}
	\begin{pgfonlayer}{nodelayer}
		\node [style=none] (0) at (-2, -2) {};
		\node [style=none] (1) at (2, -2) {};
		\node [style=none] (2) at (-2, 2) {};
		\node [style=dot] (3) at (0, 0) {};
		\node [style=none] (4) at (2, 2) {};
	\end{pgfonlayer}
	\begin{pgfonlayer}{edgelayer}
		\draw [style=scalar] (2.center) to (3);
		\draw [style=scalar] (0.center) to (3);
		\draw [style=scalar] (3) to (4.center);
		\draw [style=scalar] (3) to (1.center);
	\end{pgfonlayer}
\end{tikzpicture}}}} & $-ig_4$ & & \raisebox{-4ex}{\raisebox{0.5ex}{\scalebox{0.3}{\begin{tikzpicture}
	\begin{pgfonlayer}{nodelayer}
		\node [style=none] (0) at (-2, -2) {};
		\node [style=none] (1) at (2, -2) {};
		\node [style=none] (2) at (-2, 2) {};
		\node [style=star] (3) at (0, 0) {};
		\node [style=none] (4) at (2, 2) {};
	\end{pgfonlayer}
	\begin{pgfonlayer}{edgelayer}
		\draw [style=scalar] (2.center) to (3);
		\draw [style=scalar] (0.center) to (3);
		\draw [style=scalar] (3) to (4.center);
		\draw [style=scalar] (3) to (1.center);
	\end{pgfonlayer}
\end{tikzpicture}}}} & $- ig_4(\delta m_H^2 + 2\delta_H - 2\delta_v)$
\end{tabular}
\end{table}

We do not list the rules for the gluon self-interactions, since any diagrams involving these vertices are identically zero by colour. The analytic expressions for the counterterm insertions  $\delta_X$ are as follows:
\begin{align}
	\begin{split}
	    \delta m_t =& -\frac{g_t}{2m_t^2}\Bigl[\Bigl(\frac{g_3 m_t}{m_H^2} - g_t\Bigr)\tilde{A}_0(m_H^2) + g_t \Bigl(1 -8\frac{m_t^2}{m_H^2}N_c\Bigr)\tilde{A}_0(m_t^2)\\
    &+ g_t(m_H^2 - 4 m_t^2) \tilde{B}_0(m_t^2, m_H^2, m_t^2)\Bigr]
	\end{split}\\
	\begin{split}
	    \delta_t =& +\frac{g_t^2}{2 m_t^2} \Bigl[ \Bigl( (3 - 2 \epsilon) + 4 (\epsilon - 1) \frac{m_t^2}{m_H^2} \Bigr) \tilde{A}_0(m_H^2) + (2\epsilon - 3) \tilde{A}_0(m_t^2)\\
    &+ (2\epsilon - 3) (m_H^2 - 2 m_t^2) \tilde{B}_0(m_t^2, m_H^2, m_t^2) \Bigr]
	\end{split}\\
	\begin{split}
	    \delta m_H^2 =& -\frac{1}{2 m_H^2} \Bigl[ \Bigl( \frac{g_3^2}{m_H^2} - g_4 \Bigr) \tilde{A}_0(m_H^2) + 8 g_t N_c \bigl( g_t - g_3 \frac{m_t}{m_H^2} \bigr) \tilde{A}_0(m_t^2)\\
	&- g_3^2 \tilde{B}_0(m_H^2, m_H^2, m_H^2) - 4 g_t^2 (m_H^2-4m_t^2) N_c  \tilde{B}_0(m_H^2, m_t^2, m_t^2) \Bigr]
    \end{split}\\
	\begin{split}
	    \delta_H =& +\frac{1}{3 m_H^2}\Bigl[ \frac{g_3^2}{m_H^2} (\epsilon - 1) \tilde{A}_0(m_H^2) + 12 g_t^2 N_c (1 - \epsilon) \tilde{A}_0(m_t^2)\\
	&+ \frac{g_3^2}{2} (2 - \epsilon) \tilde{B}_0(m_H^2, m_H^2, m_H^2) \\&- 6 g_t^2 ((1 - \epsilon)m_H^2  + 2 m_t^2) N_c \tilde{B}_0(m_H^2, m_t^2, m_t^2) \Bigr]
    \end{split}\\
	\delta T =& -\frac{g_3}{2} \tilde{A}_0(m_H^2) + 4 g_t m_t N_c \tilde{A}_0(m_t^2)\\
	\delta_v =& \frac{1}{2^D \pi^{D/2}}\frac{1}{2 v^2} \left(-\frac{m_H^2}{2} + N_c m_t^2 - 2 N_c A_0(m_t^2) - 3 A_0(m_H^2) + 8 N_c \frac{m_t^2}{m_H^2} A_0(m_t^2)\right)\label{eq:dvev}
\end{align}
As explained in Appendix~\ref{ssec:dvev}, $\delta_v$ cannot be split up in different coupling structures, since we obtain the full expression from \cite{Biekotter:2023xle}, where this is not provided.\\
The scalar integrals are defined to be
\begin{align}
	\tilde{A}_0\!\left(m_1^2\right) :=&\ \frac{1}{2^D\pi^{D/2}}A_0\!\left(m_1^2\right)=\frac{\mu^{4-D}}{2^D\pi^{D/2}}\int\frac{\mathrm{d}^D\ell}{i\pi^{D/2}}\frac{1}{\ell^2-m_1^2}\\
	\begin{split}
        \tilde{B}_0\!\left(p^2,m_1^2,m_2^2\right) :=&\ \frac{1}{2^D\pi^{D/2}}B_0\!\left(p^2,m_1^2,m_2^2\right)\\=&\ \frac{\mu^{4-D}}{2^D\pi^{D/2}}\int\frac{\mathrm{d}^D\ell}{i\pi^{D/2}}\frac{1}{(\ell^2-m_1^2)((\ell+p)^2-m_2^2)}	  
	\end{split}
\end{align}
with the t'Hooft scale, $\mu$, to repair the dimensionality and the causal $i \delta$ Feynman prescription understood implicitly.

\subsection{Comparison of Counterterms and Renormalization Procedures}
\label{app:comparison}
In this section, we briefly compare the renormalisation procedure used for the vev in our work, given in Eq.~\eqref{eq:vev0} and Eq.~\eqref{eq:dvev}, to the schemes presented in Ref.~\cite{Biekotter:2023xle} and Ref.~\cite{Denner:2019vbn}.

After dropping all non-SM terms, the vev renormalization in Eq.~(2.18) of Ref.~\cite{Biekotter:2023xle} reads
\begin{equation}
    \frac{1}{v_{T,0}^2}=\frac{1}{v_\mu^2}\big[1-\frac{1}{v_\mu^2}\Delta v_\mu^{(4,1,\mu)}\big]=\frac{1}{v_\mu^2}\big[1-\frac{1}{v_\mu^2}\Delta \tilde v_\mu^{(4,1,\mu)}-\frac{1}{v_\mu^2}\Delta v_{\mu,\mathrm{tad}}^{(4,1,\mu)}\big]\,\,,
\end{equation}
where we have used Eq.~(A.14) of the same reference to collect all contributions not associated with tadpoles in $\Delta \tilde v_\mu^{(4,1,\mu)}$ and the remaining tadpole contributions in $\Delta v_{\mu,\mathrm{tad}}^{(4,1,\mu)}$.
Using the relations,
\begin{align}
v_{T,0} |_\text{\cite{Biekotter:2023xle}} &\equiv v_0 + \Delta v, \\
v_\mu |_\text{\cite{Biekotter:2023xle}} &\equiv v
\end{align}
and inserting Eq.~\eqref{eq:vev0}, we obtain,
\begin{align}
\frac{1}{v_{T,0}^2} = \frac{1}{(v_0 + \Delta v)^2} \approx \frac{1}{v^2 (1 + 2 \delta_v + 2 \frac{\Delta v}{v})} \approx \frac{1}{v^2} (1 - 2 \delta_v - 2 \frac{\Delta v}{v})
\end{align}
where, in the last two manipulations, we retain only terms linear in $\delta_v$ and $\Delta v$.
By comparison, we can identify
\begin{align}
\Delta v_\mu^{(4,1,\mu)}|_\text{\cite{Biekotter:2023xle}} & \equiv 2v^2 \big(\delta_v+\frac{\Delta v}{v}\big), \\
\Delta v_{\mu,\mathrm{tad}}^{(4,1,\mu)}|_\text{\cite{Biekotter:2023xle}} & \equiv 2v\Delta v~.
\end{align}

The comparison of our counterterms to those given in Ref.~\cite{Denner:2019vbn}, is less straightforward, as they instead use the renormalisation constants $\delta M_W^2, \delta s_w, \delta Z_e$ to parametrise the renormalisation, where $e$ is the electric charge, $s_w = \sin \theta_w$,  and $\theta_w$ is the Weinberg angle.
Using the tree level relation for the bare vev,
\begin{equation}
\frac{2 M_{W,0} s_{w,0}}{e_0} = v_0
\end{equation}
we obtain
\begin{equation}
\frac{\delta M_W^2|_\text{\cite{Denner:2019vbn}}}{2 M_W^2} + \frac{\delta s_w|_\text{\cite{Denner:2019vbn}}}{s_w} - \delta Z_e|_\text{\cite{Denner:2019vbn}} = \delta v, 
\end{equation}
where the extra factors of $M_W$ and $s_w$ in the denominator are due to their definition $M_{i,0}^2 = M_i^2 + \delta M_i^2$, for $i=W,Z$ rather than e.g. $M_{i,0}^2 = M_i^2\,(1 + \delta M_i^2)$, see Eq.~(98) of Ref.~\cite{Denner:2019vbn}.
This allows us to express our counter terms, given in Section~\ref{app:rulesExpr}, in terms of their renormalisation constants.
To match our counterterm expressions exactly, we additionally set $\delta t_\mathrm{PRTS} |_\text{\cite{Denner:2019vbn}} =0 $ and $\delta t_\mathrm{FJTS} |_\text{\cite{Denner:2019vbn}} = \delta T $ in their expressions, i.e. we select the Fleischer-Jegerlehner tadpole scheme.
Finally, to recover the counterterm insertions we give in Eq.~\eqref{eq:dvev}, the $\delta Z_e$ expression should be derived in the $G_\mu$ scheme, as described in Section~5.1.1 of Ref.~\cite{Denner:2019vbn}.

\section{Analytic Continuation of the Master Integrals}
\label{app:continuation}
In order to determine how we should perform the analytic continuation of the master integrals from the Euclidean region to the physical region, we first remark that the Feynman ``$i\delta$'' prescription in momentum space corresponds, in Feynman parameter space, to the following replacement in the second Symanzik polynomial $\mathcal{F}$: 
\begin{equation}
    \mathcal{F}\ \longrightarrow\ \mathcal{F}-i\delta
\end{equation}
where we have used the postive-definiteness of the first Symanzik polynomial $\mathcal{U}$. This prescription must be applied consistently in order to obtain valid results when crossing threshold singularities. From the diagram-constructable definition of the second Symanzik polynomial $\mathcal{F}$, this corresponds to sending all the (squared) momenta flowing between two $2$-forests, $s(T_1,T_2)$, to $s(T_1,T_2)+i\delta$  and internal masses squared, $m_{i}^{2}$, to $m_{i}^{2}-i\delta$ \cite{Hidding:2020ytt}. 

In {\sc DiffExp}, we encounter segments centred on threshold singularities and their expansions may involve multi-valued functions such as logarithms or square roots. At these points, the $\delta$-prescription supplied as an input by the user is applied. Given that $s(T_1,T_2)$ are linear combinations of the Mandelstam variables and external masses (which also obey conservation relations), {\sc DiffExp} instead takes as input a list of irreducible polynomials (``DeltaPrescriptions'') in the dimensionless variables of the problem which are zero on these threshold singularities and an additional term $\pm i\delta$ to prescribe the branch choice. For physical threshold singularities, the correct choice of $\pm i\delta$ is essential to obtain accurate results but {\sc DiffExp} also requires a choice to be made for polynomials which go to zero on non-physical singularities and this choice can be freely made without affecting results. 

In practice, similarly to \cite{PhysRevD.108.036024}, we construct the power sets of both the external momenta and the internal masses and generate a list of $\delta$-prescriptions of the form ${s-m^{2}+i\delta}$ where $s$ is a generalised squared sum of momenta and $m^{2}$ is a generalised squared sum of internal masses. We obtain a list of irreducible polynomials which are zero on singularities of our partial derivative matrices $\mathbf{A}_{x_{i}}$ and then see which correspond directly to prescriptions in our constructed list and give them the correct sign of $i\delta$ by expanding the given irreducible polynomial about that point. The remaining singularities are non-physical and we arbitrarily assign $+i\delta$. This method generates correct results for all points checked so far with \pysecdec{} and changing the sign of a prescription for a polynomial corresponding to a physical singularity can be explicitly seen to give the wrong result.

\section{Bare Form Factors}
\label{app:bareplots}
\begin{figure}[ht]
    \centering
    \begin{subfigure}{0.495\textwidth}
    \centering
    \includegraphics[width=0.6\textwidth]{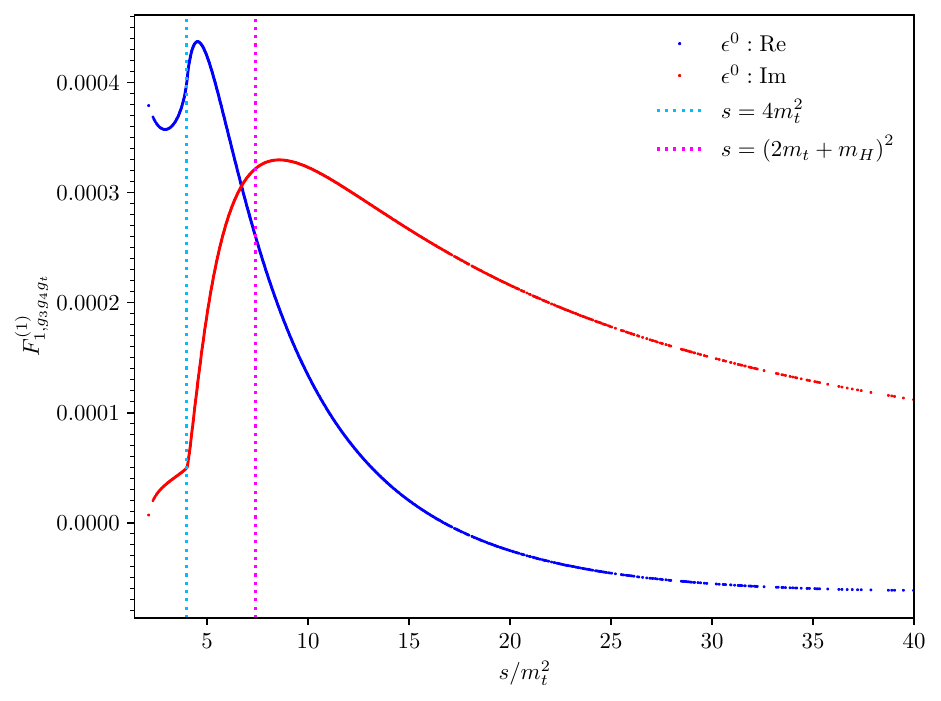}
    \caption{$F^{(1)}_{1, g_3 g_4 g_t}$}
    \end{subfigure}
    \hfill
    \begin{subfigure}{0.495\textwidth}
    \centering
    \includegraphics[width=0.6\textwidth]{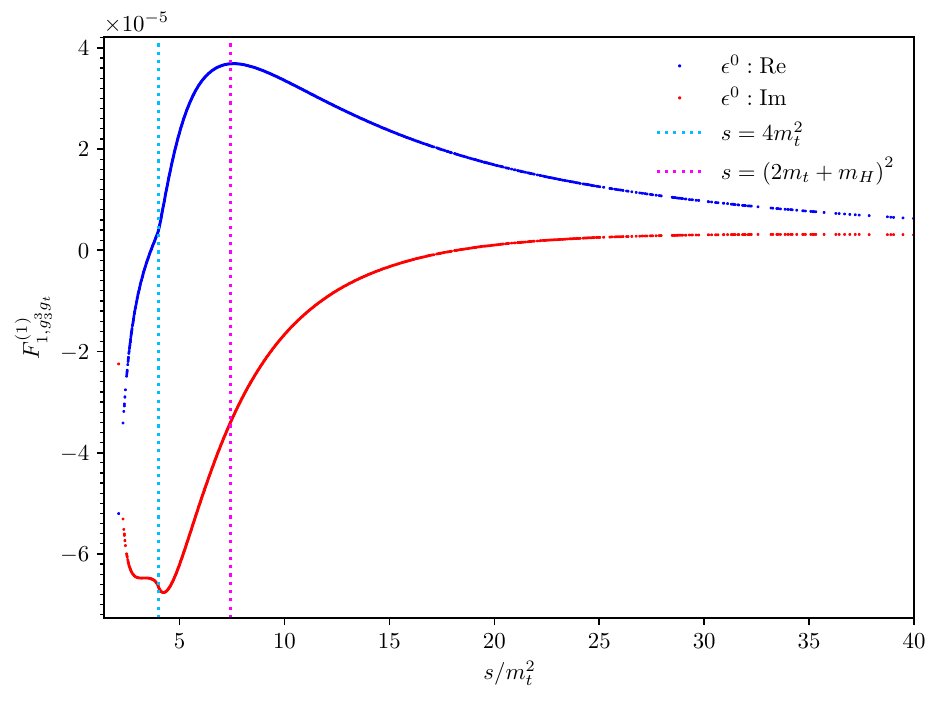}
    \caption{$F^{(1)}_{1, g_3^3 g_t}$}
    \end{subfigure}
    \\
    \centering
    \begin{subfigure}{0.495\textwidth}
    \centering
    \includegraphics[width=0.6\textwidth]{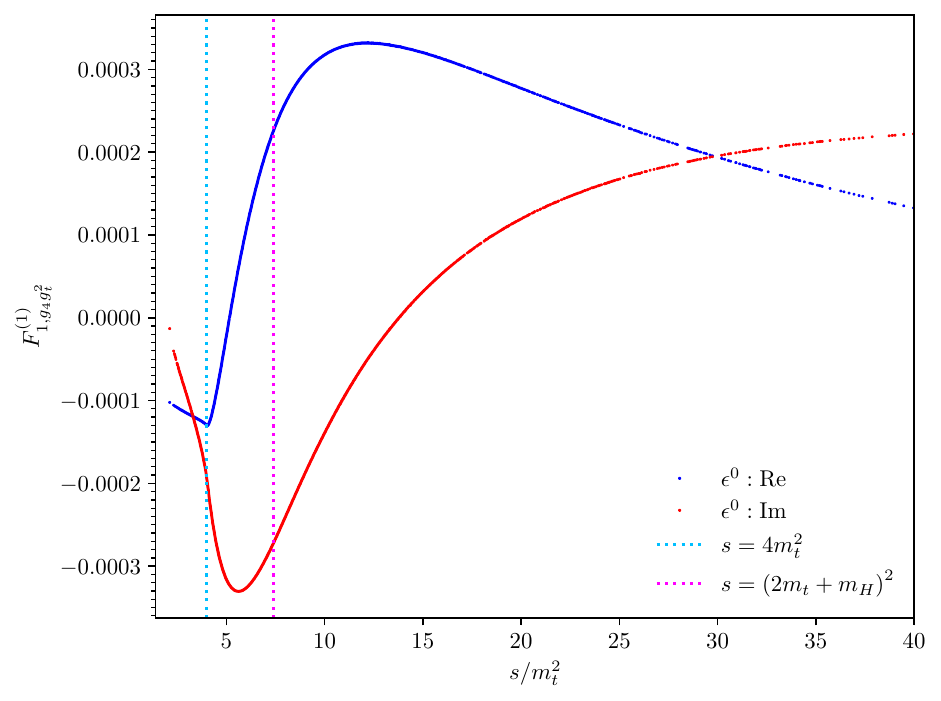}
    \caption{$F^{(1)}_{1, g_4 g_t^2}$}
    \end{subfigure}
    \\
    \begin{subfigure}{0.495\textwidth}
    \centering
    \includegraphics[width=0.6\textwidth]{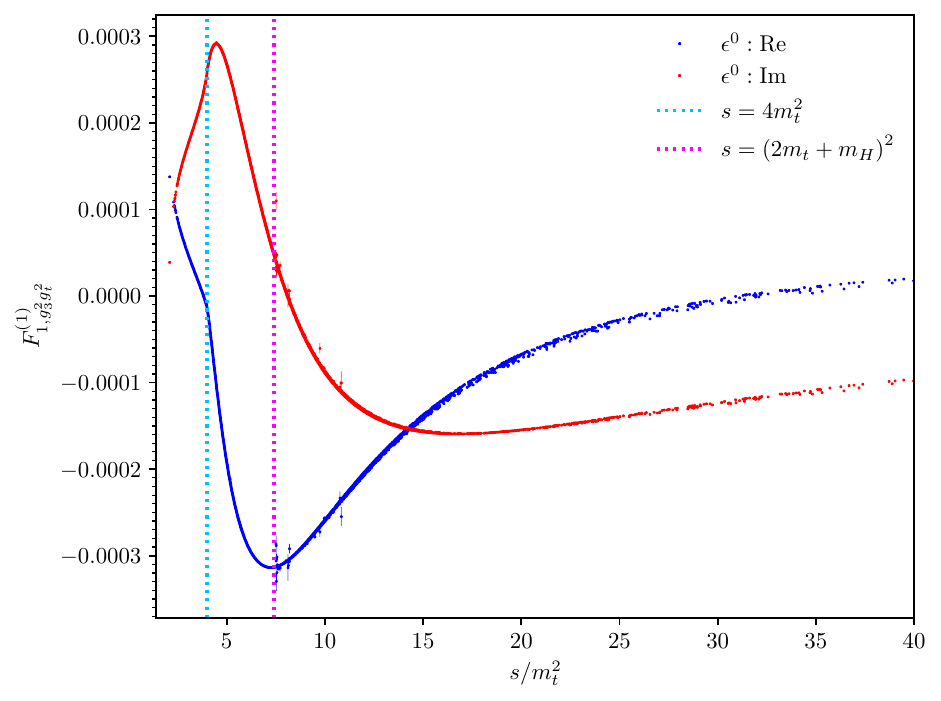}
    \caption{$F^{(1)}_{1, g_3^2 g_t^2}$}
    \end{subfigure}
    \hfill
    \centering
    \begin{subfigure}{0.495\textwidth}
    \centering
    \includegraphics[width=0.6\textwidth]{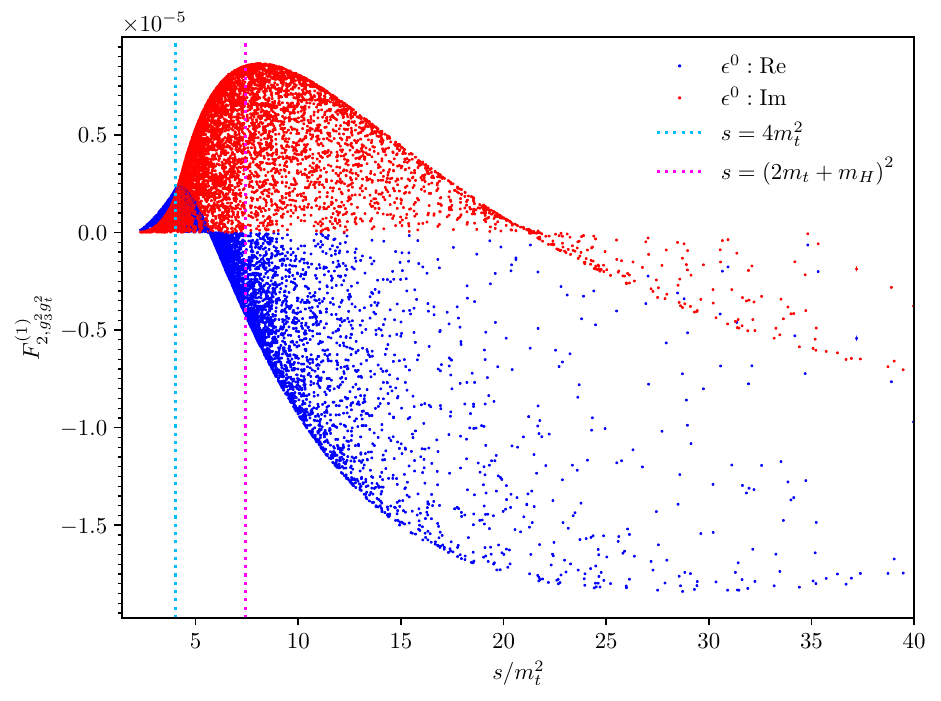}
    \caption{$F^{(1)}_{2, g_3^2 g_t^2}$}
    \end{subfigure}
    \\
    \begin{subfigure}{0.495\textwidth}
    \centering
    \includegraphics[width=0.6\textwidth]{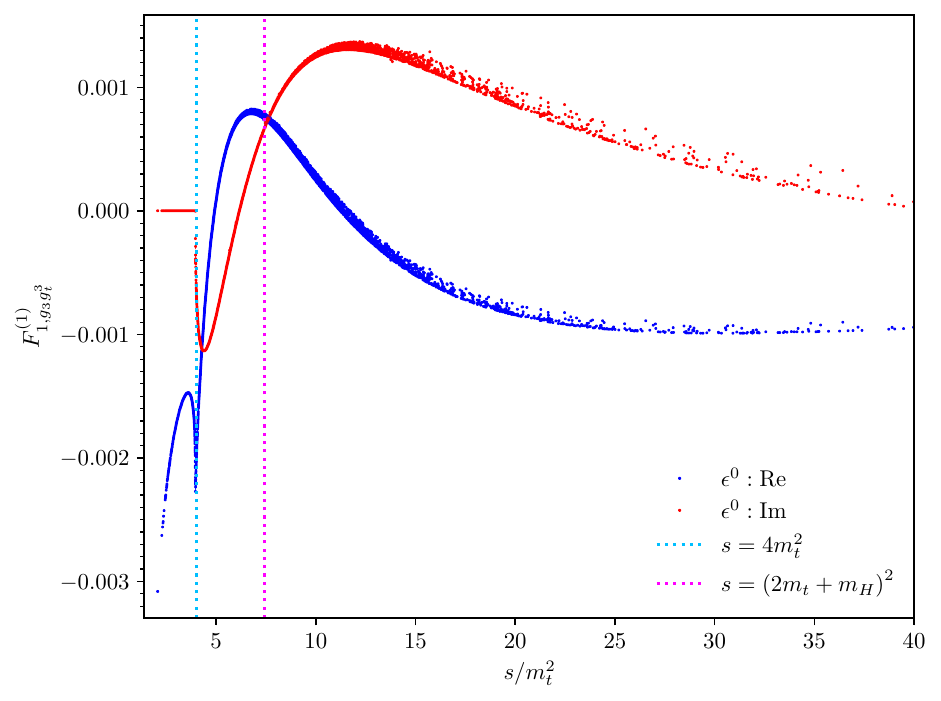}
    \caption{$F^{(1)}_{1, g_3 g_t^3}$}
    \end{subfigure}
    \hfill
    \centering
    \begin{subfigure}{0.495\textwidth}
    \centering
    \includegraphics[width=0.6\textwidth]{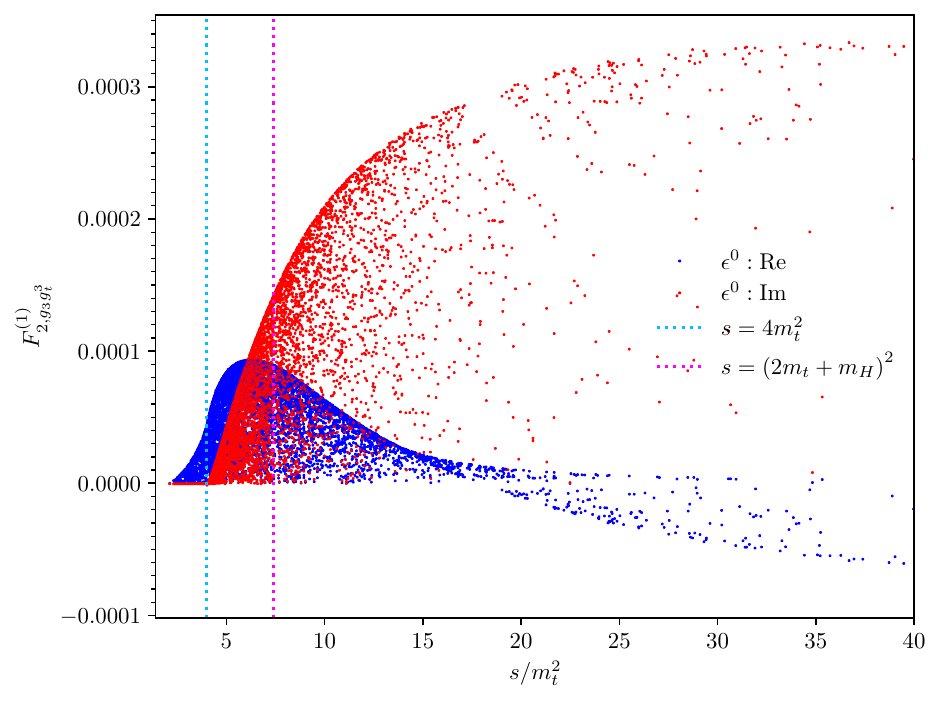}
    \caption{$F^{(1)}_{2, g_3 g_t^3}$}
    \end{subfigure}
    \\
    \begin{subfigure}{0.495\textwidth}
    \centering
    \includegraphics[width=0.6\textwidth]{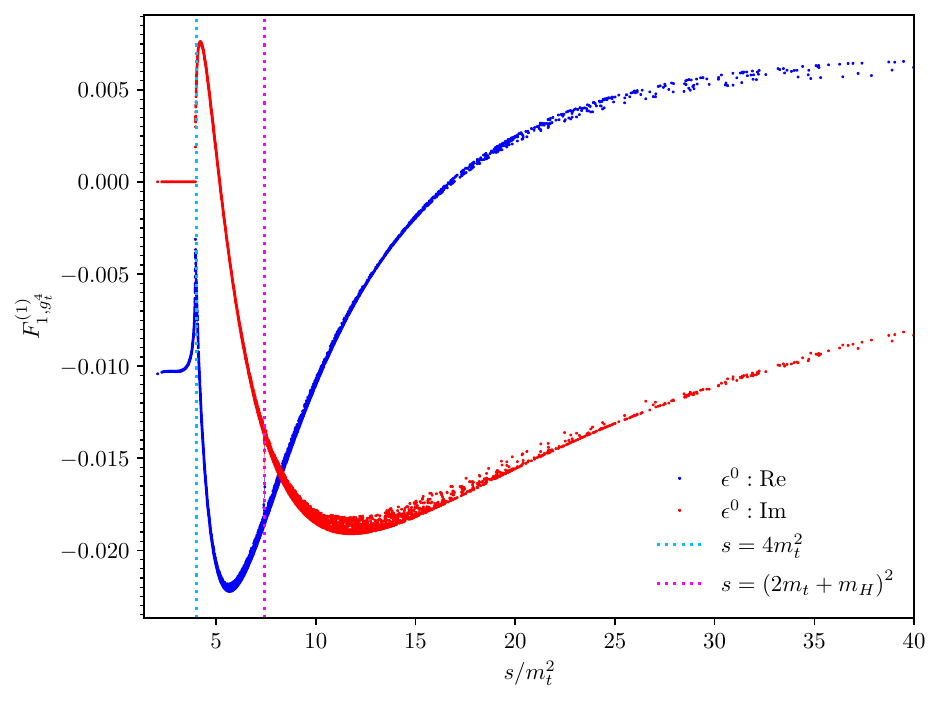}
    \caption{$F^{(1)}_{1, g_t^4}$}
    \end{subfigure}
    \hfill
    \centering
    \begin{subfigure}{0.495\textwidth}
    \centering
    \includegraphics[width=0.6\textwidth]{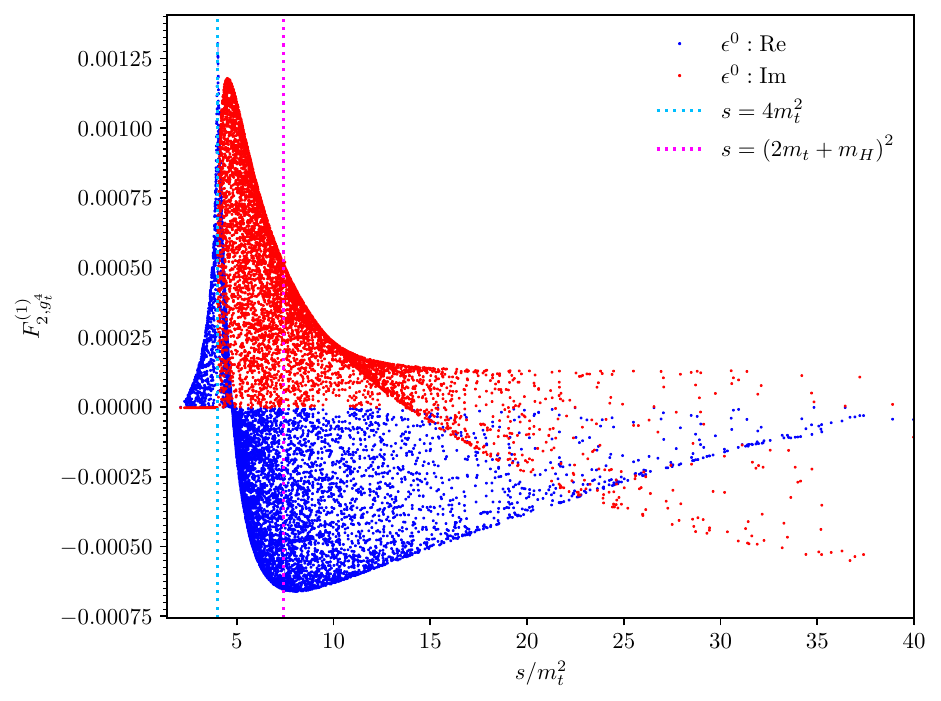}
    \caption{$F^{(1)}_{2, g_t^4}$}
    \end{subfigure}
    \caption{Plots of the $\epsilon^0$ coefficient of the bare form factors separated on coupling structure, $F^{(1)}_{i,j}$.}
    \label{fig:barestrucplots}
\end{figure}
In Fig.~\ref{fig:barestrucplots}, we plot the bare form factors separated on coupling structures as in Eq.~\eqref{eq:couplingStructures}. The spread of points, which is due to $t$-dependence, is more pronounced in the contributions to $F_2^{(1)}$ than in the contributions to $F_1^{(1)}$.

\clearpage

\bibliographystyle{JHEP}
\bibliography{main}

\end{document}